\documentclass[preprint,showpacs,preprintnumbers,amsmath,amssymb,nofootinbib]{revtex4}

\usepackage{etex}

\usepackage{amssymb,amsthm,amscd,amsbsy,array}
\usepackage{bm}
\usepackage{soul} 

\usepackage{graphics,graphicx,xcolor}

\usepackage{soul} 

\usepackage[colorlinks=true, pdfstartview=FitV, linkcolor=blue, citecolor=blue, urlcolor=blue]{hyperref} 

\newcommand{\gb}{\quad\colorbox{green}}

\newenvironment{redtext}{\color{red}}
{\ignorespacesafterend}
\newenvironment{bluetext}{\color{blue}}{\ignorespacesafterend}

\newenvironment{magentatext}{\color{magenta}}{\ignorespacesafterend}
\newenvironment{cyantext}{\color{cyan}}{\ignorespacesafterend}
\newenvironment{orangetext}{\color{orange}}
{\ignorespacesafterend}

\newcommand{\bmagenta}{\begin{magentatext}}
\newcommand{\emagenta}{\end{magentatext}}
\newcommand{\bcyan}{\begin{cyantext}}
\newcommand{\ecyan}{\end{cyantext}}
\newcommand{\bblue}{\begin{bluetext}}
\newcommand{\eblue}{\end{bluetext}}
\newcommand{\bred}{\begin{redtext}}
\newcommand{\ered}{\end{redtext}}
\newcommand{\borange}{\begin{orangetext}}
\newcommand{\eorange}{\end{orangetext}}

\numberwithin{equation}{section}
\renewcommand{\theequation}{\thesection.\arabic{equation}}
\let\ssection=\section
\renewcommand{\section}{\setcounter{equation}{0}\ssection}
\newcommand{\beq}{\begin{equation}}
\newcommand{\eeq}{\end{equation}}

\newcommand{\cA}{{\mathcal{A}}}

\newcommand{\bb}{{\bf b}}

\newcommand{\bbeta}{\boldsymbol{\beta}}

\def\where{{\quad\text{where}\quad}}

\newcommand{\diag}{\mathrm{diag}}

\newcommand{\cL}{{\mathcal{L}}}

\newcommand{\bp}{{\bf p}}

\newcommand{\bx}{{\bm{x}}}

\renewcommand{\Re}{\mathrm{Re}}

\newcommand{\SO}{\mathrm{SO}}

\newcommand{\bX}{{\bf X}}

\newcommand{\dx}{\dot{x}}

\def\smallover#1/#2{\hbox{$\textstyle\frac{#1}{#2}$}} %

\def\bp{{\bm{p}}}

\def\parag{\hfil\break} 
\def\kikezd{\parag\underbar}

\def\bequ{\begin{enumerate}}
\def\eequ{\end{enumerate}}
\def\bitem{\begin{itemize}}
\def\eitem{\end{itemize}}

\def\beq{\begin{equation}}
\def\eeq{\end{equation}}
\def\beqa{\begin{eqnarray}}
\def\eeqa{\end{eqnarray}}
\def\nn{\nonumber}
\def\barray{\left(\begin{array}}
\def\earray{\end{array}\right)}
\def\barraynb{\begin{array}}
\def\earraynb{\end{array}}

\def\IR{{\mathbb{R}}} 

\def\?{{\quad\gb{\fbox{\texttt{?}}\;}}\quad}
\def\p{{\partial}}

\def\v0{\mathbf{0}}

\def\p{\partial}

\def \p{{\partial}}

\def\be {{\bf e}}

\def\6{\partial}
\def\7{\tilde}
\def\8{\widehat}
 \def\bx{{\bf x}}

\def\pa{\partial}

\def\G11{\Gamma_{11} }

\newcommand{\const}{\mathop{\rm const.}\nolimits}
\newcommand{\half }{\smallover{1}/{2}}

\def\smallover#1/#2{\hbox{$\textstyle\frac{#1}{#2}$}} %
\def\smallcirc{{\raise 0.5pt \hbox{$\scriptstyle\circ$}}}
\def\2{{\smallover1/2}}

\renewcommand{\theequation}{\thesection.\arabic{equation}}
\let\ssection=\section
\renewcommand{\section}{\setcounter{equation}{0}\ssection}

\def\besub{\begin{subequations}}
\def\esub{\end{subequations}}

\begin{document}


\title{Lukash plane waves, revisited}

\author{
M. Elbistan$^{1,2}$\footnote{mailto:Mahmut.Elbistan@lmpt.univ-tours.fr},
P. M. Zhang$^{1}$\footnote{corresponding author. mailto:zhangpm5@mail.sysu.edu.cn},
G. W. Gibbons$^{3}$\footnote{
mailto:G.W.Gibbons@damtp.cam.ac.uk},
P. A. Horvathy$^{4}$\footnote{mailto:horvathy@lmpt.univ-tours.fr}
}

\affiliation{
${}^1$ School of Physics and Astronomy, Sun Yat-sen University, Zhuhai, (China)
\\
${}^2$ Physics Department,
Bo\u{g}azi\c{c}i University,
34342 Bebek / Istanbul, (Turkey)
\\
${}^3$ D.A.M.T.P., Cambridge University, U.K.
\\ Wilberforce Road,
\\ Cambridge CB3 0WA, (U.K.)
\\
${}^4$  Institut Denis-Poisson CNRS/UMR 7013 - Universit\'e de Tours - Universit\'e d'Orl\'eans Parc de Grammont, 37200; Tours, (France)
}

\date{\today}


\begin{abstract}
The Lukash metric is a homogeneous gravitational wave which at late times approximates the behaviour of a generic class of spatially homogenous cosmological models with monotonically decreasing energy density. The transcription from Brinkmann to
Baldwin-Jeffery-Rosen (BJR) to Bianchi coordinates is presented and the relation to a  Sturm-Liouville equation is explained. The 6-parameter isometry group is derived. In the Bianchi VII range of parameters we have two BJR transciptions. However using either of them induces a mere relabeling of the geodesics and isometries. Following pioneering work of Siklos, we provide a self-contained account of the geometry and global structure of the spacetime. The latter contains a Killing horizon to the future of which the spacetime resembles  an anisotropic version of the Milne cosmology and to the past of which it resemble the Rindler wedge.
\bigskip

JCAP \textbf{01} (2021), 052
doi:10.1088/1475-7516/2021/01/052
[arXiv:2008.07801 [gr-qc]].
\end{abstract}

\maketitle

\tableofcontents

\section{Introduction}\label{Intro}

The standard approach to cosmology is to assume the
\emph{Cosmological Principle} which says
that  the Universe and its matter content are spatially
homogeneous and isotropic ($SO(3)$ invariant) \cite{Hawking:1973uf,Ehlers,exactsol}.
This leads, locally at least, to the Friedmann-Lema\^{\i}tre-Robertson-Walker (FLRW) metric
\beq
ds^2 = - dt^2 + a^2(t)\,  d\Omega^2_K \,,
\eeq
where $d\Omega^2_K$ is the metric of constant  curvature $K$
on
\begin{itemize}
\item  3-sphere \;\,\qquad \quad $ \mathbb{S}^3= SO(4)/SO(3)$  \quad \, if\; $K=+1$\,,

\item Euclidean space \,$ \mathbb{E}^3  =  E(3) / SO(3)$ \,\,\,\,\,\quad if\;\, $K=0 $\,,

\item  hyperbolic space $ \mathbb{H}^3 = SO(3,1)/SO(3)$  if\; $K=-1$\,.
\end{itemize}
For  general $a(t)$ the continuous isometries
 above, which act on the  hypersurfaces of constant cosmic time $t$, are maximal.
  However if $a(t) = e^{Ht} $ and $K=0$
where $H=\sqrt{ \frac{3}{\Lambda}}$, or if $K=-1$ and
$a(t) = \frac{1}{H} \sinh (Ht)$, for example, then, despite appearances, the continuous isometries
are much larger: $SO(4,1)$.

Typically  the metric is singular at times when $a(t)=0$.
The singularity is fictitious because the  FLRW coordinates break down at $t=-\infty$ or $t=0$, respectively.
The full spacetime accessible by past-directed timelike geodesics of finite propertime is
 de Sitter  spacetime  $SO(4,1)/SO(3,1)$  which is homogeneous \cite{Hawking:1973uf}.

 An even more striking example is obtained by setting  $K=-1$ and  take the limit $ H \downarrow 0$. Then
 we find that $a(t) =t$, which is the Milne metric and is in fact flat. The FLRW coordinates cover only the interior of the future
light cone of  a point in Minkowski spacetime $E(4,1)/SO(3,1)$. Evidently past-directed time-like geodesics
can leave the interior of the light cone in finite propertime. It is also clear that the same phenomenon
occurs  when $H \ne 0$.

A natural question to ask is whether such a behavior persists in the case
of more general anisotropic  cosmological models, i.e., those admitting a  three-dimensional Bianchi-type  subgroup $G_3$
of continuous isometries acting on spacelike hypersurfaces but for which  there is no $SO(3)$ subgroup fixing  points on those hypersurfaces.

The aim of the present paper is to explore a class of Ricci-flat  solutions of the Einstein equations which exhibit a  structure which  is very similar to  the Milne case~: these are the \emph{Lukash solutions} \cite{Lukash,Lukash76,Lukash76,exactsol}, which are of Bianchi type $VII_h$.
This case is generic among Bianchi type groups, because it  depends upon a dimensionless parameter $h$. The full isometry group is six-dimensional and admits a four dimensional subgroup which acts transitively on the complete spacetime.

Henceforth we restrict our attention at the $VII_h$ case. Other possibilities will be studied elsewhere.

Two Ricci-flat pp-waves  are known to admit
a six dimensional, multiply transitive isometry group \cite{Ehlers,exactsol,Sippel}.
They are~: the
``anti-Mach metric'' \cite{OS} which is a circularly polarised periodic (CPP) plane wave  \cite{exactsol},
and the Lukash plane wave \cite{Lukash,Lukash74,Lukash76}.

CPP   is a non-singular continuous  gravitational wave \cite{OS,exactsol,Sippel,CPP}. Historically, it provided a powerful argument for the physical existence  in Einstein's theory of gravitational waves in the
complete absence of material sources or localised patches of curvature, since  the metric is spacetime-homogeneous, i.e.,
invariant under the action of a four-dimensional simply-transitive group of symmetries.

The metric of the Lukash  wave (to which this paper is devoted)  may be cast in Brinkmann coordinates,
\beq
ds^2 = 2 dU dV + d{\bX}^2 + K(U,\bX)\, dU^2\,
\label{BrinkmannMetric}
\eeq
with profile
\beq
K = -  2\Re \bigl(C \zeta^2 U^{2(i\kappa-1)} \bigr)\,
\label{Lukashprofile}
\eeq
where  $\zeta=({X^1}+i{X^2})/\sqrt{2}$ \cite{Siklos91}.
Physically, $C$ represents the strength (amplitude) of the wave and  $\kappa$ represents the polarisation. Although both could be real and $\kappa$ could even be complex, we shall only consider $C$ and $\kappa$ both positive.

 Brinkman coordinates  are well-defined for all $V,\bX$ and $U>0$, but they break down at $U=0$.
The nature  of the singularity at $U=0$ was the subject of
a number of investigations \cite{Siklos81,Siklos2,Siklos3,Siklos91}
subsequent to \cite{Collins:1972tf,Collins:1973lda,Lukash,Lukash74,Lukash76} and motivated in part by \cite{EllisKing1}.

Following \cite{Siklos91} the
 Bianchi VII group structure requires that the parameters satisfy \cite{Siklos91},
\beq
\text{\small either}\quad
0 \leq C < \kappa
\quad\text{\small or} \quad
 \kappa  = C\, > 1/2\,
\label{BVIIrange}
\eeq
we shall refer to as the \emph{Bianchi VII range}.

The Lukash solutions, which  contain the Milne metric as a special case,  have   attracted considerable attention in the past  because of their cosmological applications. They have been shown to approximate at late times a wide class of \emph{spatially homogeneous} (or {cohomogeneity one})
cosmological models with vanishing or negligible
cosmological constant, in which the matter density also becomes  negligible at late times \cite{Collins:1972tf,Collins:1973lda,EllisKing1,Barrow86,Barrow:2004gk,Fliche,Fliche2,Fliche3}.

A striking result of \cite{Collins:1972tf,Collins:1973lda}
is that they do \emph{not} isotropise, i.e., they do \emph{not} approximate the Milne metric at late times. In fact they are stable at late times
\cite {Barrow86,Barrow:2004gk,Wainwright:1998ms,Hsu}. Thus, contrarily  to what had been believed
previously, they did  not provide a natural  answer to the question~: \emph{``Why is the Universe Isotropic ?~''} raised in ref. \cite{Collins:1973lda}.

General accounts of anisotropic cosmological  models may be found in \cite{exactsol,EllisMacCallum,MacCallum}.

The  stability  may be partially understood from the other reason for which the Lukash metrics have attracted interest~:
they are a special example of a \emph{plane gravitational wave}~: they are exact solutions of the vacuum Einstein equations which generically
have a five dimensional  isometry group $G_5$ \cite{exactsol,BoPiRo}, which acts multiply  transitively on three dimensional null hypersurfaces identified as  the wave fronts.

 $G_5$ is conveniently found by switching to another coordinates system attributed to Baldwin, Jeffery and Rosen (BJR) in which the metric has the form
\beq
ds^2 = 2du dv + a_{ij}(u) dx^i dx^j\,.
\label{BJRmetric}
\eeq
Then it is found \cite{Sou73,Carroll4GW,Carrollvs} that
$G_5$ is subgroup of L\'evy-Leblond's six dimensional Carroll group \cite{LL}.
The relation between Brinkmann and BJR coordinates entails solving a matrix-valued Sturm-Liouville equation \cite{SL-C}.

The isometry group of Lukash plane waves has (as all pp-waves do) a three dimensional abelian subgroup
consisting of translations of the coordinates $v,x^i$. This subgroup acts  simply transitively  on the wave fronts,
hence the appelation ``plane''.  The coordinates constructed in \cite{Siklos81,Siklos2,Siklos3,Siklos91}
are in fact a set of BJR coordinates, in which the Lukash plane wave is  manifestly  plane symmetric.
 This isometry also renders  the geodesic equations integrable. This fact had played a role in the memory effect for plane  gravitational waves \cite{Memory}.

For Bianchi type cosmological models, the temporal coordinate is
typically chosen as proper time $\tau$
along the orthogonal  trajectories  of those orbits.
If matter is present in the form of a perfect fluid,
these orthogonal trajectories may coincide with the fluid flow lines as happened in the case of  Friedmann-Lema\^{\i}tre models. If the fluid flow lines are not orthogonal to the  orbits of $G_3$ then the model is referred to as ``tilted'' \cite{EllisKing2}.

In \cite{Siklos81,Siklos2,Siklos3,Siklos91} the author

\goodbreak
\begin{itemize}
\item
Constructed a set of un-tilted  coordinates for Lukash
 plane waves adapted to the Bianchi  type $VII_h$ group:

\item Showed that these coordinates break down at finite
 comoving time in the past at a fictitious singularity at which the orbits become lightlike. In other words their is a Killing horizon ; before that time the orbits are timelike.
\end{itemize}
These results  are scattered
among four papers published over a number of years.
One of the intentions of the present paper is to provide
a systematic and self-contained derivation in a uniform notation and conventions, consistent with current  work on gravitational waves.

\smallskip
The organisation of the paper is as follows.
In section \ref{LukashSec} we  cast the Brinkmann form of the Lukash metric first into BJR, and then to Bianchi form.

The isometries are determined in sec.\ref{Isosection}.
An important aspect  is that it reveals, \emph{within a suitable range of parameters},  \eqref{BVIIrange},
the existence  of a three dimensional subgroup of the six dimensional isometry which is of Bianchi $VII_h$ type. This group acts transitively  on three dimensional orbits and leads
to an intimate connection between the theory of gravitational waves and that of spatially homogeneous cosmological models and thence to the theory of Killing horizons. Since these topics are not necessarily familiar to researchers in gravitational waves we have provided  a brief overview in an Appendix.

In the range
$ 0 < C < \kappa$
we get two different transcriptions from Brinkmann
 to BJR coordinates see sec. \ref{Multis}, which lead to
two types of  $VII_h$ groups and thus two different foliations.
 The one of interest for making the connection with the work of
\cite{Collins:1972tf,Collins:1973lda,Lukash,Lukash74,Lukash76} is spacelike and only covers part of the spacetime.
The two different transcriptions induce two sets of geodesics and isometries, related by an inversion of the light-cone coordinate,
\beq
u \to \frac{1}{u}\,,
\label{univ}
\eeq
which plainly interchanges $u=0$ and $u=\infty$.
Section \ref{ExampleSec} illustrates our theory on
 examples.

In  section \ref{GlobalSec} we provide a global picture of spacetime. The gravitational wave emanates from a singular wave front and is divided by a Killing horizon into two regions which we have dubbed  of \emph{Milne type} and of \emph{Rindler type}.

In the  Milne region the orbits of the Bianchi
group are spacelike and the spacetime resembles an anisotropic deformation of Milne's cosmological model. In the Rindler region the orbits are timelike and the spacetime resembles an anisotropic deformation of the Rindler wedge.

\section{Lukash Plane waves: from Brinkmann to BJR to Bianchi}\label{LukashSec}

The aim of this section is to show that the Lukash metric \eqref{BrinkmannMetric}-\eqref{Lukashprofile} may  locally be cast first into the BJR form \eqref{BJRmetric}, and then into that of a spatially homogeneous metric
\beq
ds^2 =-dt^2 + g_{ij}(t) \lambda^i \lambda ^j \,, \label{Bianchiform}
\eeq
where $\lambda^i $ are left-invariant one forms on a group
of Bianchi type $VII_h$ (see the Appendix A).

\subsection{From Brinkmann to BJR~: Siklos' theorem}\label{SiklosSec}

We start with the Lukash metric \cite{Lukash,Lukash74,Lukash76,exactsol,Terzis:2008ev} written in
Brinkmann coordinates. Eqn.  $\#$ (3.1) of \cite{Siklos91} adapted to our conventions is,
\beq
ds^2 = 2dU dV + 2 d \zeta d\bar{\zeta} - C \big(U^{2(i\kappa -1)}\zeta^2 + U^{-2(i \kappa+1)}\bar{\zeta}^2\big)dU^2\, .
\label{3.1}
\eeq
  This metric depends on two real parameters $C\geq0$ and $\kappa>0$~: $C$ determines the strength of the wave and $\kappa$ its frequency.
The sign of $\kappa$ fixes also the sense of the polarization ;
$\kappa>0$  will be chosen in what follows.
We note for further reference that when $U > 0$ then  $U^{\kappa}=e^{\kappa\ln U}$ allows us to present the profile of  \eqref{3.1} in a real form,
\beq
 - \frac{C}{U^2} \Big[\cos\big(2\kappa\ln(U)\big)\big(({X^1})^2-({X^2})^2\big) - 2\sin\big(2\kappa\ln(U)\big)
 {X^1}{X^2} \Big]dU^2\,.
\label{realBLukash}
\eeq

We now state :

\vskip-3mm
\kikezd{Theorem (Siklos) \cite{Siklos91}}~:
\textit{The coordinate transformation $(U,\zeta,V) \to (u,\xi,v)$ defined by
\besub
\begin{align}
\label{un1}
\zeta &= e^{i\alpha} u^{s-i\kappa} \big[ \xi u^{ib}\cosh(\mu/2) - \bar{\xi} u^{-ib}\sinh(\mu/2) \big]\, ,
\\[3pt]
V &= v -\frac{u^{2s -1}}{2}\left[2s\, \xi\bar{\xi} \cosh(\mu) - \big((s+ib) \xi^2 u^{2ib} + c.c.  \big)\sinh(\mu)  \right]\, ,
\label{un2}
\end{align}
\label{un12}
\esub
augmented with $U = u$ carries the Brinkmann-form metric \eqref{3.1} to}
\beq
ds^2 = 2du dv + u^{2s} \left[2 \cosh(\mu) d\xi d\bar{\xi} -\sinh(\mu)\big(u^{2iks} d\xi^2 + u^{-2iks} d\bar{\xi}^2 \big) \right] \,,
\label{A.1}
\eeq
where the parameters satisfy a series of constraints \cite{Siklos91},
\besub
\begin{align}
b &= ks\,,
\\
b\cosh(\mu) &=\kappa\, ,
\\
 s -s^2 &= \kappa^2 \tanh^2(\mu)\, ,
 \label{c2r}
 \\
 C\cos(2\alpha) &=-2\kappa^2\tanh(\mu) \, ,
 \\
C\sin(2\alpha) &=(2s-1) \kappa\tanh(\mu)\, .
\end{align}
\label{Sconstraints}
\esub
The proof is obtained by a term-by-term calculation \cite{Siklos91}.
\goodbreak

Eliminating the auxiliary variables $\mu$ and $\alpha$ yields
four ({\rm a priori} complex) solutions,
\beq
s = \dfrac {1}{2}\pm \dfrac{1}{\sqrt{2}}
 \sqrt{\smallover1/4-\kappa^2
\pm \sqrt{(\smallover1/4+\kappa^2)^2- C^2}}\,;
\label{4s}
\eeq
assuming $s >0$, the other parameters are expressed as,
\beq\begin{array}{clllc}
k &=& \displaystyle\sqrt{\frac{\kappa^2 + s^2-s}{s^2}}\,,
&\qquad\quad &b=ks=\sqrt{\kappa^2 + s^2-s},
\\[8pt]
\sinh \mu &=& \sqrt{\displaystyle\frac{s(1-s)}{\kappa^2 + s^2-s}}\,,
&\quad &\tan(2\alpha) = \dfrac{1/2-s}{\kappa}\quad\,.\;
\end{array}
\label{kbmualpha}
\eeq
\goodbreak
Putting  $\xi=x^1+ix^2$  shows, moreover, that  \eqref{A.1} is  of the BJR form  \eqref{BJRmetric}  with profile matrix
$a=(a_{ij})$ whose entries are \vspace{-2mm}
\beq
\begin{array}{cll}
a_{11} &=& u^{2s} \left[\cosh(\mu)-\sinh(\mu) \cos\big(2b\ln(u)\big) \right]\, ,
 \\
a_{12}=a_{21}&=&  u^{2s} \sinh(\mu)\sin\big(2b\ln(u) \big)\, ,
\\
a_{22}&=& u^{2s} \left[\cosh(\mu) + \sinh(\mu)  \cos\big(2b\ln(u)\big)\right]\, .
\end{array}
\label{BJRa}
\eeq
The $\ln u$ here clearly requires $u > 0$.

The metric is thus decomposed into the sum of ($u$-dependent)
 background plus  perturbation terms,
\beq
2du dv+u^{2s}\cosh\mu\; d\bx\cdot{  \barray{ccc}1& &0\\ 0 &&1\earray}d\bx
\;-\;
u^{2s}\sinh\mu\,d\bx\cdot
\barray{cc}
\cos (2b\ln u)& - \sin (2 b \ln u) \\
- \sin (2 b\ln u) & - \cos (2b \ln u)
\earray d\bx
\,,
\label{LukBJRprofile}
\eeq
When $\mu \neq 0$, putting $\ell=-\coth\mu$ we can present \eqref{LukBJRprofile}  in a form considered in \cite{Siklos81,Siklos91},
\beq
ds^2=\;2du dv -
\frac{u^{2s}}{\sqrt{\ell^2-1}}
d\bx\cdot\barray{cc}
\ell+\cos (2b \ln u)& - \sin (2 b \ln u)  \\
- \sin (2 b\ln u) & \ell- \cos (2b \ln u)
\earray\,{d\bx}\,.
\label{Sik81form}
\eeq

The exponent $s$ in \eqref{4s} may take multiple real values, implying  multiple transcriptions.
 The question will be further analysed in secs. \ref{Multis}, \ref{ExampleSec} and \ref{GlobalSec}.

\subsection{From BJR to Bianchi VIIh  form}\label{BianchiformSec}

We now cast  the Lukash metric (\ref{Sik81form})   into Bianchi $VII_h$ form. To this end we introduce new coordinates $t, z$ in the region for which $u>0$ and $v<0$ by setting
\beq
 u = t \exp(-\frac{z}{2b}\,) \,, \qquad v  =- \half   t \exp(\,\frac{z}{2b}\,)  \,.
\label{tzdefinition}
\eeq
We have
\beq
 2 du dv =- dt ^2 + \frac{1}{4b^2} t^2 dz^2\,,
\qquad
2b \ln u = 2b \ln t - z  \,,
\qquad u^{2s} =
t^{2s} e^{-\frac{sz}{b} } \,.
\nn
\eeq
Expanding the trigonometric functions the metric may be
cast in a form similar to that in \cite{Siklos81} sec.4.
To show that it is of the Bianchi $VII_h$  form \eqref{Bianchiform} one may use the expressions for the
left-invariant one forms $\lambda^i$ written in
 \eqref{oneforms}.  This entails
relating $\lambda$ to $z$ and $x,y$ to $\mu ,\nu$.
Then a careful examination of the term $a_{ij}dx^i dx^j$  shows that this metric can be brought to the generic Bianchi $VII_h$ form (\ref{Bianchiform}) with the identifications
\beq
\alpha = x\,, \quad \beta = y\,,\quad  \gamma = {z}/{2}\,,\quad
 c = {s}/{b}=1/k\,.
\eeq
The non-vanishing metric elements are \vspace{-2mm}
\begin{subequations}
\begin{align}
g_{11} & = t^{2s} \Big( \cosh(\mu) - \sinh(\mu)\cos(2b\ln t)  \Big)\,,
\\
g_{22} & = t^{2s} \Big( \cosh(\mu) + \sinh(\mu)  \cos(2b\ln t)  \Big)\, ,\\
g_{12} & = t^{2s} \sinh(\mu) \sin(2b\ln t)\, ,
\\
g_{33} & = {t^2}/{b^2}\, .
\end{align}
\label{gij}
\end{subequations}
Using $b=ks$
the group parameter $ h = c^{2}$ becomes \cite{Siklos81}, sec.4.1
\beq
h^{-1}= k^2 \,.
\label{Bparameter}
\eeq

We also remark that the Lukash metric is an example of a \emph{self-similar spatially homogeneous cosmology} \cite{Hsu}  since it admits a one parameter group of homotheties
\beq
(u, x^i,v,) \rightarrow (u,\lambda x^i, \lambda^2 v) \,,\quad \lambda>0\,,
\label{homothety}
\eeq
under which the metric scales as
$
ds^2 \rightarrow \lambda^2 ds^2
$  \cite{Sippel,exactsol,Conf4GW}\,.

\subsection{Relation to Sturm-Liouville}\label{BtoBJRSec}

Now we put these results into a broader perspective.
Let us recall that  the coordinate transformation \eqref{un12} carries the metric written in Brinkmann coordinates,  \eqref{BrinkmannMetric}, to the BJR form \eqref{BJRmetric}
whose profile is  \eqref{BJRa}. The  transformation \eqref{un12} fits  into the  framework of ref. \cite{Gibbons:1975jb,Deser}~:
\beq
X^i = P_{ij} x^j\,, \quad U = u\,, \quad V = v - \frac{1}{4} x^i a_{ij}^{\prime} x^j\,,
\where a_{ij}(u)=P^TP\,,
\label{BfromBJR}
\eeq
where the prime denotes $d/du$ and the matrix $P(u)$ satisfies a $2\times 2$ \emph{matrix Sturm - Liouville equation} \cite{SL-C},
\beq
\label{SLeqn}
P''_{ij} = K_{ik}P_{kj}\,,  \qquad (P^T)' P = P^T P'\,.
\eeq
Instead of solving our S-L eqn directly (which is an arduous task), we first extract the ``square-root'' matrix $P=(P_{ij})$ from \eqref{BJRa},
as,
%
\begin{subequations}
\begin{align}
P_{11} &= \frac{U^s}{2} \left[ e^{i\alpha} U^{-i\kappa} \big(U^{ib} \cosh(\mu/2) - U^{-ib}\sinh(\mu/2) \big) + c.c.   \right]\, , \\
P_{12} &= \frac{U^s}{2} \left[i e^{i\alpha} U^{-i\kappa} \big(U^{ib} \cosh(\mu/2) + U^{-ib}\sinh(\mu/2) \big) + c.c.   \right]\, , \\
P_{21} &= \frac{U^s}{2} \left[ -ie^{i\alpha} U^{-i\kappa} \big(U^{ib} \cosh(\mu/2) - U^{-ib}\sinh(\mu/2) \big) + c.c.   \right]\, , \\
P_{22} &= \frac{U^s}{2} \left[ e^{i\alpha} U^{-i\kappa} \big(U^{ib} \cosh(\mu/2) + U^{-ib}\sinh(\mu/2) \big) + c.c.   \right]\, .
\end{align}
\label{complexP}
\end{subequations}
Then a tedious calculation allows us to verify that \eqref{complexP} is indeed a solution of the Sturm~-~Liouville equation \eqref{SLeqn} for any real $s$  in \eqref{4s} and parameters given in \eqref{kbmualpha}.
 Let us emphasise that to any ``good'' choice of $s$ [i.e., such that $s$ is real] is associated a B $\to$ BJR transcription.
Illustrative examples will be studied in sec.\ref{ExampleSec}.

\section{Geodesics and isometries of the Lukash metric}\label{Isosection}

\subsection{In Brinkmann coordinates}\label{IsoB}

Let us consider a pp wave with metric $g_{\mu\nu}dx^{\mu}dx^{\nu}$  written in Brinkmann coordinates $x^{\mu}=U,X^i,V$  as in \eqref{BrinkmannMetric}, with profile
\beq
K_{ij}(U){X^i}{X^j}=
\half{\cA}_+(U)\Big((X^1)^2-(X^2)^2\Big)+{\mathcal{A}}_{\times}(U)\,X^1X^2\,,
\label{Bprofile}
\eeq
where $\cA_+$ and ${\mathcal{A}}_{\times}$ are the $+$ and $\times$ polarization-state amplitudes \cite{BoPiRo,Ehlers,exactsol}.
The geodesics 
of \eqref{BrinkmannMetric}-\eqref{Bprofile} may be obtained from the Lagrangian
$
\cL=
 \frac{1}{2} g_{\mu\nu}\dx^{\mu}\dx^{\nu}\,
$  where $\dx^{\mu}=\frac{dx^{\mu}}{d\lambda\;}$,  $\lambda$ being an affine parameter.
Independence of $\lambda$ implies that along a geodesic
\beq
\cL  =\const = - \half m^2 \, ;
\label{relmass}
\eeq
we identify $m$ with the relativistic mass.
In fact $\lambda$ is an affine parameter~: varying $\cL$ w.r.t. to $V$ implies that
$
\frac{dU}{d\lambda} = \const$, that is,  $\lambda = aU +b$. Henceforth we choose $\lambda= U$.

Varying $\cL$ w.r.t. to $\bX$ gives two linear, generally coupled, second-order equations for the transverse coordinate $\bX$ with $U$-dependent coefficients,
\beq
(X^i)'' = K_{ij}(U) X^j,
\quad\text{i.e.}\quad
\dfrac{d^2\bX}{dU^2} - \frac{1}{2}\barray{lr}
{\cA}_+ &{\mathcal{A}}_{\times}
\\
{\mathcal{A}}_{\times} & -{\cA}_+
\earray
\bX = 0\,.
\label{ABXeq}
\eeq
Notice that these equations involve only the transverse coordinate $\bX$ and are independent of the mass $m$ in
\eqref{relmass}. This equation describes the motion of a non-relativistic system with ``time'' $U$, namely an oscillator with $U$-dependent coefficients in the transverse plane \cite{Bargmann}.

Returning to the 4D relativistic system, we just mention for completeness that
varying $\cL$ w.r.t. to $V$ gives a  second order equation for $V$,
\beq
V(U)'' +\frac{1}{2} K_{ij}' X^i X^j +2 K_{ij} (X^i)' X^j=0\,,
\label{Vpp}
\eeq
which can be integrated along a solution of \eqref{ABXeq} once the latter has been found \cite{Conf4GW}.

\kikezd{Theorem} \cite{Torre,SL-C}~:
\textit{For the vacuum pp wave profile \eqref{BrinkmannMetric}  with $\Delta K= 0$, the Killing vectors are
\beq
\widehat{\beta}=\bbeta\,\p_{\bX}- \beta_i'(U) X^i\,\p_V\,,
\label{betalift}
\eeq
where the two-vector $\bbeta=(\beta_i)$ satisfies the vectorial Sturm-Liouville equation \cite{Torre,SL-C}
\beq
\label{betaSL}
\beta_i''(U) = K_{ij}(U) \beta_j(U).
\eeq
Remarkably, eqn. \eqref{betaSL} is \underline{identical} to the transverse equations of motion, \eqref{ABXeq} when $\bX$ is replaced by $\bbeta$} \footnote{We note for completeness that
$\beta_V= - \beta_i'(U) X^i\,$ cf. \eqref{betalift} implies
\beq
\beta_V'' = - \Big(K_{ij}' \beta_j + K_{ij}\beta_j'\Big) X^i\,,
\label{betaVeq}
\eeq
which is similar to but different from \eqref{Vpp} under the replacement $\bX \to \bbeta$. Thus the lifts of isometries resp. of geodesics to 4D are different.
}.

Thus problem boils down to solving \eqref{betaSL}, which admits a 4-parameter family of solutions.
Putting  $\beta = \frac{\beta_1 + i\beta_2}{\sqrt{2}} $ and $\zeta = \frac{X^1 + iX^2}{\sqrt{2}}$,
 eqns. \eqref{betalift}-\eqref{betaSL} are, in the Lukash case, \vskip-8mm
\besub
\begin{align}
&\hat{\beta} =
 \beta\partial_\zeta + \bar{\beta}\partial_{\bar{\zeta}}  - \left(\bar{\beta}^{\prime}\,\zeta +  \beta^{\prime}\,\bar{\zeta}\right)\partial_V\,,
 \\
&\beta'' + C\, U^{-2(i\kappa +1)}\, \bar{\beta} = 0\,.
\label{Lukash4Kvs}
\end{align}
\esub
\goodbreak
Solutions can be found only by a case-by-case study, see the examples in sec. \ref{ExampleSec}.

The Lukash wave admits an additional $6th$ isometry \cite{exactsol}.
 For $C=0$ UV boosts,
\beq
U   \rightarrow  e^{\tau} U\,,
\qquad
\zeta \rightarrow  \zeta\,,
\qquad
V  \rightarrow e^{-\tau} V\,,
\qquad(\tau\in\IR)
\label{UVboost}
\eeq
are isometries of the Minkowski metric $2d\bar{\zeta}d\zeta+2dUdV$, but for the Lukash metric
\eqref{3.1} they are manifestly broken by the last term. An isometry can however be obtained by combining it with another broken generator, namely
with that of transverse rotations,
\beq
U   \rightarrow   e^{\tau} U\,,
\qquad
\zeta \rightarrow  e^{i(-\kappa\tau)} \zeta\,,
\qquad
V  \rightarrow e^{-\tau} V\,,
\label{finitekUVboost}
\eeq
which leaves the Lukash metric
\eqref{BrinkmannMetric}-\eqref{Lukashprofile} (unlike the wavefront $U=\const$) invariant.
The isometry \eqref{finitekUVboost} is generated by \footnote{The construction is reminiscent of the one we observed for the Bogoslovsky-Finsler model \cite{BogoF}, where the UV-boost symmetry of Very Special Relativity \cite{GibbonsVSR} is broken but can be restored by combining it with an (equally broken)  dilation. Eqns \eqref{finitekUVboost} - \eqref{kUVboost} are actually identical to those valid for a CPP wave with $\kappa=\omega/2$, cf. \cite{Carroll4GW,CPP}.}
\beq
Y_{\kappa} =  (U\partial_U-V\partial_V)
- \kappa\, (X^1\partial_2 - X^2 \partial_1)  \,.
\label{kUVboost}
\eeq
This generator is in fact ``chronoprojective'' as defined in \cite{5chrono,Conf4GW}: it only preserves the \emph{direction} of $Y_V\equiv \p_V$ but not $\p_V$ itself~ :
\beq
L_{Y_{\kappa}}\p_V=\psi\,\p_V,
\qquad\psi=1.
\label{chronop}
\eeq

\subsection{Carroll symmetry in BJR pulled back to Brinkmann}\label{IsoBJR}

To find the isometry group in Brinkmann coordinates requires solving a S-L equation and  therefore can be dealt with only by a case-by-case study. There is however another approach, though, which uses BJR coordinates \cite{Sou73,Carroll4GW}.
 The general chrono-projective vector field is \cite{Conf4GW},
\beq
Y = Y^u (x,u) \pa_u + Y^i(x,u) \pa_{i} + \big(b(x,u) - \psi\, v\big)\, \pa_v \,,
\label{ChronoBJR}
\eeq
where
the  Killing equations require, \vspace{-3mm}
\besub
\begin{align}
&\partial_iY^u = 0\,,
\label{Yia}
\\
&\partial_u Y^u = \psi\,,
\label{Yib}
\\
&\p_uY^v = 0\,,
\label{Yic}
\\
&\partial_i Y^v +(\partial_u Y^j)a_{ij}=0\,,
\label{Yid}
\\
&Y^u (\partial_u a_{ij}) + a_{kj}(u)\partial_i \big(Y^k(x,u)\big)+ a_{ki}\big(\partial_j Y^k(x,u)\big) = 0 \,.
\label{Yie}
\end{align}
\label{Yiall}
\esub
When $\psi=0$ i.e.,  when the transformation leaves $\p_v$ invariant, we end up with the $5$ parameter broken Carroll algebra \cite{Carroll4GW,Conf4GW} ,
\beq
\label{aCarroll}
Y_{Carr} = h \partial_v + c^i \partial_i +  b_i (S^{ij}\partial_j - x^i\partial_v)\,,
\qquad
S^{ij}(u) = \int^u\!\!a^{ij} (\tilde{u}) d\tilde{u}\,,
\eeq
where $h, c^i, b_i$ are constants and  $(a^{ij}) = a^{-1}$ denotes the inverse  matrix of the BJR profile $a=(a_{ij})$.
$S=S^{ij}(u)$ is the Souriau  matrix \cite{Sou73,Carroll4GW}.

Three translations (one ``vertical" and two transverse) with parameters $h$  and  $f_i$, respectively, are read off immediately.
The last term in $Y_{Carr}$ (we call Carroll boosts) requires to calculate the Souriau matrix. Assuming that $s\neq\2$ we find,
\begin{subequations}
\begin{align}
\hskip-5mm
S^{11}
&=u^{1-2s}\left(\frac{\cosh \mu}{1-2s}+
\frac{\sinh\mu}{4b^{2}+\left(2s-1\right)^{2}}
\Big[(1-2s)\cos (2b\ln u)+2b\sin (2b\ln u) \Big]\right)\,
\\[8pt]
\hskip-5mm
S^{12} &= S^{21}
= u^{1-2s}\left(\frac{\sinh \mu }{4b^{2}+\left(2s-1\right)^{2}}\Big[2b\cos(2b\ln u)+\left( 2s-1\right) \sin(2b\ln u)\Big]\right)\,
\\[8pt]
\hskip-5mm
S^{22}
&= u^{1-2s}\left(\frac{\cosh \mu}{1-2s}-
\frac{\sinh\mu}{4b^{2}+\left(2s-1\right)^{2}}\Big[(1-2s)\cos (2b\ln u) +
2b\sin(2b\ln u)\Big]\right)\,.
\end{align}
\label{Smatexp}
\end{subequations}
For $s=\half$ see \eqref{Ck1Smatrices}.
\goodbreak

The 6th isometry which has $\psi=1$ (cf. \eqref{chronop})  will be dealt with in the next subsection.

In conclusion, we obtain the BJR form of  4 isometry vectors which, in the flat case would reduce to translations and Galilei boosts in transverse space, see sec.\ref{ExampleSec} below.

Once the isometries have been identified in BJR coordinates, we can pull them back  to Brinkmann by inverting \eqref{BfromBJR}.
Applied to \eqref{aCarroll} a lengthy calculation  yields,
\beqa
Y_{Carr} &=& h \frac{\partial}{\partial V} + c^i \left(P_{ji} \frac{\partial}{\partial X^j} - (P_{ji})' X^j  \frac{\partial}{\partial V}\right)
+ b_i\left( S^{ik}P_{jk} \frac{\partial}{\partial X^j} - (S^{ik}P_{jk})' X^j  \frac{\partial}{\partial V} \right) \qquad \qquad
\label{addBrCa}
\eeqa
where $h, c^i, b_i$ are  the same (translation and boost) parameters as in (\ref{aCarroll}).  The elements of the S-L and Souriau matrices $P$ and $S$ are given in (\ref{complexP}) and (\ref{Smatexp}), respectively. They are worked out in   Appendix B.
\goodbreak

To sum up, a convenient way to find all symmetries in Brinkmann coordinates is to  :
\bequ
\item
Switch first to BJR coordinates following Siklos's theorem (sec. \ref{SiklosSec}, \eqref{un12}), or equivalently, through the $P$ matrix \eqref{BfromBJR}-\eqref{SLeqn} in sec \ref{BtoBJRSec}).

\item
In BJR coordinates the isometries are given by \eqref{aCarroll}.

\item
At last, we pull them back to B-coordinates using \eqref{BfromBJR} to yield  \eqref{addBrCa}.
\eequ

The main difficulty is to find the $P$-matrix, which requires to solve the SL equation \eqref{SLeqn}, -- a task for which we  have no general method. However in the Bianchi VII case \eqref{BVIIrange} the Siklos transcription discussed in section \ref{SiklosSec} does provide us with the P-matrix  \eqref{complexP} which can then be used whenever the parameters  in \eqref{Sconstraints} and in \eqref{4s} - \eqref{kbmualpha} are all real.

\subsection{The 6th isometry and the Bianchi algebra}\label{uvbianchi}

Siklos \cite{Siklos91} shows that \eqref{Bianchiform} is the most general Bianchi VII metric;
the
Bianchi VII algebra is implemented on the $t=\const$ hypersurface with $x,y,z$ are coordinates by the generators,
\beq
\label{BianchiVIIKvs}
Z_1 = \frac{\partial }{\partial x},
\qquad
Z_2 = \frac{\partial }{\partial y}\,,
\qquad
Z_6 = 2\frac{\partial }{\partial z} + (x-ky) \frac{\partial }{\partial x} + (y+kx) \frac{\partial }{\partial y}\,\,,
\eeq
where
$
k^2 = h^{-1}
$
cf. \eqref{Bparameter}
 characterizes the sub-cases. Let us recall (sec.\ref{BianchiformSec}) that the coordinates used here are related to those in BJR according to  \eqref{tzdefinition}.

Now we confirm the above result.
$Z_1$ and $Z_2$ are plainly those translations in \eqref{aCarroll}.
Moreover, starting  with \eqref{ChronoBJR} another lengthy calculation  shows that the \emph{triple combination} of an \emph{$u-v$ boost} with a \emph{transverse rotation} and a \emph{transverse dilation}  (all of them broken individually) written in BJR coordinates,
\beq
Y_6 =
(u\partial_u-v\partial_v) - b(x\partial_y - y\partial_x) - s\,(x\partial_x+y\p_y) \,
\label{Y6}
\eeq
does indeed satisfy the constraints \eqref{Yiall}.
The generator $Y_6$ does not belong to the Carroll algebra, though, only to its 1-parameter ``chrono-Carroll'' extension
 \cite{5chrono,Conf4GW}~: it  preserves the direction of $\p_v$ only, $L_{Y_{\kappa}}\p_v=\p_v $, cf. \eqref{chronop}.

The vector fields  in \eqref{BianchiVIIKvs} generate
the Bianchi group ${G}_3$, composed of two transverse translations plus the Very Special Relativity type \cite{GibbonsVSR} triple combination \eqref{Y6}.

To conclude this subsection, we observe that~:
\bequ
\item
Pushing forward $Y_6$ in \eqref{Y6} to Brinkmann coordinates the 6th Killing vector $Y_{\kappa}$ is recovered.

\item
The BJR $\to$ Bianchi  coordinate change \eqref{tzdefinition}  in sec.\ref{BianchiformSec} carries \eqref{Y6}  to $-b\,Z_6$ in \eqref{BianchiVIIKvs}.
\eequ

\section{Multiple B $\rightarrow$ BJR transcriptions}\label{Multis}


When the SL equation \eqref{SLeqn} has several real solutions, implying multiple B $\to$ BJR transcriptions. This is what happens
in the interior of the Bianchi VII range \eqref{BVIIrange}, i.e., when
\beq
0 < C < \kappa\,.
\label{BVIIrangeInt}
\eeq
Then $s$ in \eqref{4s} is real when the plus sign is chosen in front of the inner root, yielding \emph{two real values for $s$},
\beq
s_\pm = \frac{1}{2}\pm\Delta\,,
\qquad
\Delta = \dfrac{1}{\sqrt{2}}
 \sqrt{\smallover{1}/{4}-\kappa^2
+ \sqrt{(\smallover{1}/{4}+\kappa^2)^2- C^2}}\, .
\label{C<ks12}
\eeq
Then our road map  above would provide us, {\rm a priori}, not with 4, but with 8 ``translation-boost type'' isometries, -- 4 for each BJR implementations -- which would contradict the general statement.
This does \emph{not} happen, though, as we show it now.
First we notice that
\beq
s_+  + s_-  = 1,
\quad
b_{+}=b_{-},\quad \mu_{+}=\mu_{-},\quad \alpha_{+}=-\alpha_{-}.
\label{+-parameters}
\end{equation}
Moreover, introducing
$
u_{+}=\frac{1\;}{u_{-}}
$
we deduce from eqns \eqref{complexP} and \eqref{BJRa} that the $P$-matrix and the BJR profile  behave under
$u_+ \to u_-$ and $s_+ \to s_-$
 as \goodbreak
\besub
\begin{align}
&\left\{\begin{array}{cll}
P_{11}^{+}\left(u_{+}\right)
&=&
\displaystyle\frac{1\;}{u_{-}}\,P_{11}^{-}\left(u_{-},s_{-}\right)
\\
P_{12}^{+}\left(u_{+}\right)
&=&
-\displaystyle\frac{1}{u_{-}}\,P_{12}^{-}(u_{-})
\\
P_{21}^{+}\left(u_{+}\right)
&=&
-\displaystyle\frac{1}{u_{-}}\,P_{21}^{-}(u_{-})
\\
P_{22}^{+}\left(u_{+}\right)
&=& \displaystyle\frac{1\;}{u_{-}}\,P_{22}^{-}\left(
u_{-}\right)
\end{array}\right.\,,
\label{Ps+s-}
\\[8pt]
&\left\{\begin{array}{cll}
a_{11}^{+}\left(u_{+}\right)
&=&\displaystyle\frac{1\;}{u_{-}^{2}}\,a_{11}^{-}\left(u_{-}\right)
\\
a_{12}^{+}\left(u_{+}\right) = a_{21}^{+}\left(u_{+}\right)
&=&-\displaystyle\frac{1\;}{u_{-}^{2}}\,a_{12}^{-}\left(u_{-}\right)=-\displaystyle\frac{1\;}{u_{-}^{2}}\,a_{21}^{-}
\\
a_{22}^{+}\left(u_{+}\right)
&=&
\displaystyle\frac{1\;}{u_{-}^{2}}\,a_{22}^{-}\left(u_{-}\right)
\end{array}\right.\,.
\label{as+s-}
\end{align}
\label{PA+-}
\esub

Note the minus signs in the off-diagonal terms.
From this we infer the Souriau matrix
\begin{equation}
\begin{array}{cll}
S^{11}_{+}\left(u_{+}\right) &=&-S^{11}_{-}\left(u_{-}\right)
\\
S^{12}_{+}\left(u_{+}\right) &=&\quad S^{12}_{-}\left(u_{-}\right)
\\
S^{22}_{+}\left(u_{+}\right) &=&-S^{22}_{-}\left(u_{-}\right)
\end{array}\quad .
\label{+-Smatrix}
\end{equation}

\begin{figure}[ht]
\hskip-2mm
\includegraphics[scale=.3]{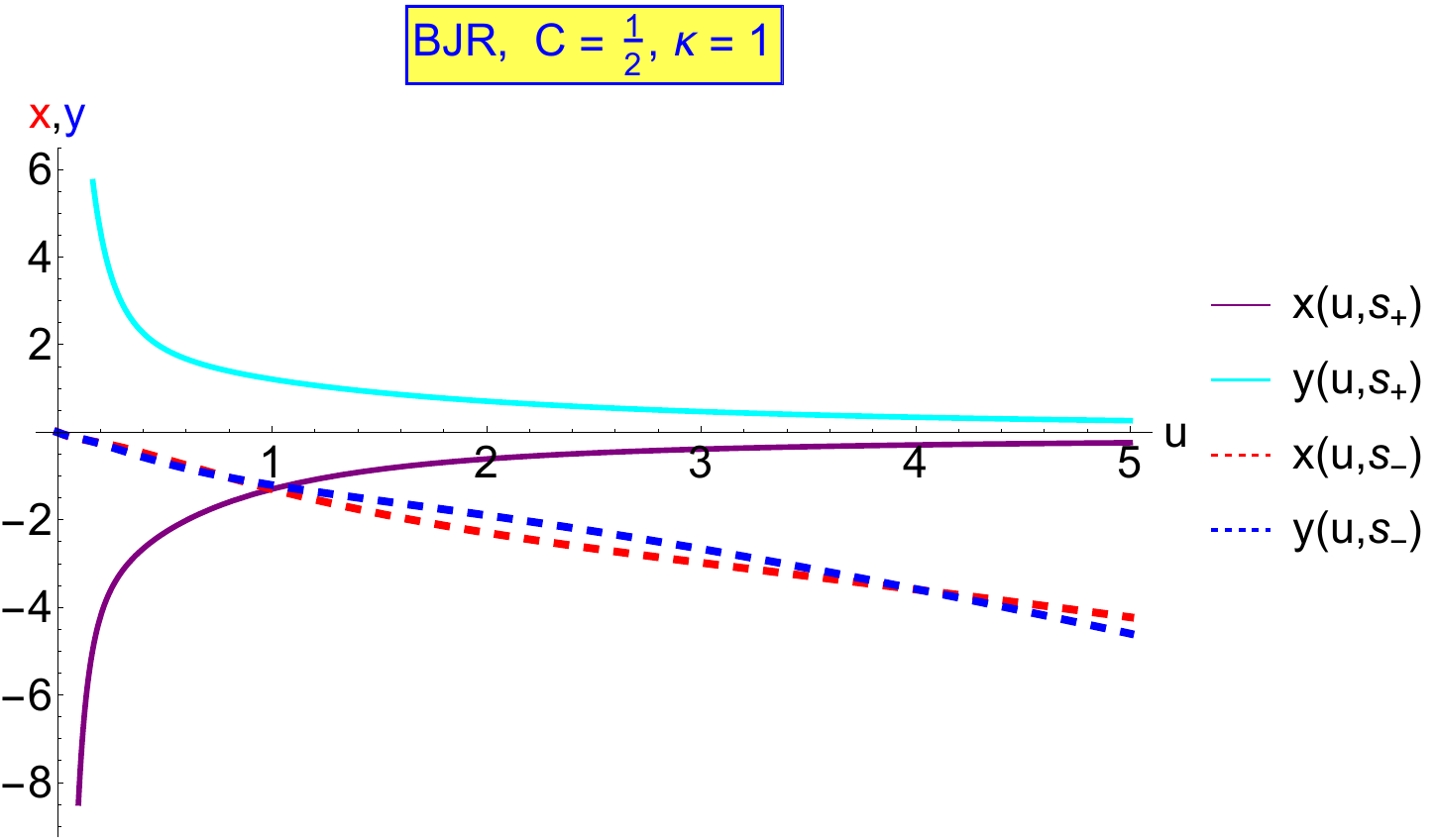}
\vskip-5mm
\caption{\textit{\small The projections to the transverse  plane  of the geodesics in BJR coordinates, unfolded to ``time" $u$.
 The trajectories obtained by choosing $s_+$ or $s_-$ look substantially different.
}
\label{fig1}}
\end{figure}

\kikezd{In BJR  coordinates}. 
%
Let us first recall that  in BJR coordinates the geodesics  are \cite{Carroll4GW},
\besub
\begin{align}
x^{i}(u)&= \; S^{ij}(u)p_{j}+x_{0}^{i}\,,
\\
v(u)&=-\frac{1}{2}\bp\cdot S(u)\bp +eu +v_0\,,
\end{align}
\label{BJR0geo}
\esub
where $p_j,\, x_0^i, e$ and $v_0$ are integration constants  determined by the initial conditions. The constant of the motion  $e=\half g_{\mu\nu}\dot{x}^\mu\dot{x}^\nu$ here is minus half of the mass square  of the geodesic, cf. \eqref{relmass}. Thus $e=0$ for a null geodesic.
Using \eqref{PA+-} and \eqref{+-Smatrix} we then find that
the transverse components behave under
$u_+ \to u_-$ and $s_+ \to s_-$ as,
\besub
\begin{align}
x_{+}\left(u_{+},
 p_{x},p_{y},x_{0}\right) &=\;\; x_{-}\left(
u_{-},-p_{x},p_{y},x_{0}\right) \,,
\label{+-BJRx}
\\
y_{+}\left(u_{+}, p_{x}, p_{y},y_{0}\right) &=-y_{-}\left(
u_{-},-p_{x},p_{y},-y_{0}\right)\,,
\label{+-BJRy}
\\
v_{+}\left(u_{+}, p_{x}, p_{y},v_0\right)
&=
-v_{-}\left(u_{-},-p_{x}, p_{y},-v_0\right)\,.
\label{+-BJRv}
\end{align}
\label{+-BJRxyv}
\esub
where the suffices $\pm$ refer to $s_{\pm}$.
Note the minus sign in front of $y$ (but not $x$) and for the integration constants.
The transverse relations \eqref{+-BJRx}-\eqref{+-BJRy} are  valid for any $e \propto m^2$, but  \eqref{+-BJRv} holds  only for $m=0$.
In conclusion, the null geodesics of the two BJR metrics are interchanged under
\beq
u_+ \to u_-\,,
\;
p_x \to - p_x\,,  \; p_y \to p_y\,, \;
x_0 \to  x_0\,,\; y_0 \to -y_0,\; v_0 \to -v_0\,;
\; s_+ \to s_-\,,
\label{+-uxyv}
\eeq
as illustrated by figs.\ref{fig1} and \ref{fig2} in sec.\ref{C1k2subsub}.

The relations between the geodesics come in fact  from one between the two BJR metrics \eqref{BJRmetric} associated with $s_\pm$. These metrics are conformally related,
\beqa
ds_{BJR,+}^{2}\left(u_{+}\right)
&\equiv& 2du_{+}dv_{+}+a_{ij}^{+}(u_{+})dx^i_+ dx^j_+
\nn\\
&=&
\frac{1}{u_{-}^{2}}\,\Big(2du_{-}dv_{-}+a_{ij}^{-}
(u_{-})dx^i_- dx^j_-\Big)
\equiv
\frac{1}{u_{-}^{2}} \,ds_{BJR,-}^{2}\left(u_{-}\right)\,\;
\label{confBJR+-}
\eeqa
under the mapping
\begin{equation}
u_{+}=\frac{1\;}{u_{-}}\,, \text{ \ \ }%
x_{+}=x_{-}\,,\text{ \ \ }y_{+}=-y_{-}\,,\text{ \ \ }v_{+}=-v_{-}
\label{BJRs+s-}
\end{equation}
 as seen by using \eqref{BJRa} and \eqref{+-parameters}.
This mapping interchanges the $v<0$ and $v> 0$ regions which in sec.\ref{GlobalSec} will be called Milne and Rindler regions.

\kikezd{In Brinkmann coordinates}. 
%
Pulling back the BJR expressions \eqref{BJR0geo} obtained for $s_\pm$ to Brinkmann coordinates using \eqref{BfromBJR} and  \eqref{Ps+s-} we get
\besub
\begin{align}
X_{+}^1\left(U_{+},p_{x},p_{y},x_{0}\right)
&=\;\;\; U_{+}X_{-}^1\left(\frac{1}{U_{+}}%
,-p_{x},p_{y},x_{0}\right)\,,
\label{+-BX}
\\
 X_{+}^2\left(U_{+},p_{x},p_{y},y_{0}\right)
&=-U_{+}X_{-}^2\left(\frac{1}{U_{+}},-p_{x},p_{y},-y_{0}\right)\,,
\label{+-BY}
\\
V_{+}(U_{+}, p_{x},p_{y},x_{0},y_{0}, v_0) & = - V_{-}(\frac{1}{U_{+}}, -p_{x},p_{y},x_{0},-y_{0}, - v_0) \nn
\\
&\quad\, -\dfrac{1}{2}U_{+}\left(\bX_{-}(\frac{1}{U_{+}}, -p_{x},p_{y},x_{0},-y_{0},)\right)^2\,,
\end{align}
\label{Bgeo+-}
\esub
where $U_{\pm}=u_{\pm}$. The last relation here is valid  for massless geodesics.
Comparison with \eqref{+-BJRv} shows that, unlike for BJR coordinates, the vertical Brinkmann coordinate $V$ does \emph{not simply change sign}~: it  has an additional term
$
-\dfrac{1}{2}U_{+}\left(\bX_{-}(\frac{1}{U_{+}})\right)^2
=
-\dfrac{1}{2U_{+}}\left(\bX_{+}(U_{+})\right)^2.
$

\goodbreak
Spelling out the formulas in Appendix B a lengthy calculation to shows that replacing $s_+$ by $s_-$ carries the $s_+$-boosts $Y_3^+$ and $Y_4^+$ (resp. translations) into a linear combination of $s_-$-translations
$Y_1^+$ and $Y_2^+$ (resp.boosts). The explicit formulas are not inspiring and are therefore omitted.

\goodbreak
\section{Examples}\label{ExampleSec}

Further insight can be gained by looking at simple examples.

\subsection{The Minkowski resp. Milne case}\label{Minkowskicase}

Before studying Bianchi VII type situations, it is instructive to look at the Minkowski case $C=\kappa=0$.   The general real solution of the S-L eqn \eqref{SLeqn} is
\beq
P(u)= \diag(A_{--}, A_{-+})
+
u\,\diag(A_{+-}, A_{++}) \,,
\label{MinkP}
\eeq
where the $A_i$ are $4$ real constants.
The splitting  of the P-matrix yields  two BJR transcriptions which correspond to two types of foliations of flat spacetime~:
\begin{itemize}
\item
Choosing
$A_{--}=A_{-+}=1$  but
$A_{+-}= A_{++}=0$  we get the  standard light-cone Minkowski form \vspace{-7mm}
\besub
\begin{align}
u_-&=U,\quad \bx_-=\bX,\quad  v_-=V\,,
\\
P_{ij}^{-}&=a_{ij}^{-} = \delta_{ij}, \quad S^{ij}_-=u_-\,\delta^{ij}
\\
ds^2_-&=2du_-dv_-+d\bx_-^2\,.
\end{align}
\label{Minkform}
\esub
The Brinkmann and BJR coordinates coincide and eqn \eqref{addBrCa} allows us to recover the 4-parameter subalgebra of the Poincar\'e algebra generated by the $4$-parameter vectorfield
\beq
Y_{B} \equiv Y^{-}_{BJR} = c^i \frac{\partial}{\partial X^i}
+ b_i\left( U\frac{\partial}{\partial X^i} - X^i  \frac{\partial}{\partial V} \right)\,.
\label{addBrCaM}
\eeq
 These generators correspond to translations and Galilei boosts in the transverse space lifted to 4D Bargmann space
\footnote{The Minkowski space can be viewed as a  of limit when the amplitude $C \downarrow 0$. Then the above results correspond  formally to  putting  $s,\, \mu \to 0$ in \eqref{4s} - \eqref{kbmualpha}.}. The geodesics are, according to \eqref{BJR0geo}
\beq
\bx_-(u_-)=\bp\,{u_-}+\bx_0, \qquad v_-(u_-)= -\half \bp^2\,{u_-}+v_0\,,
\label{s0BJRgeo}
\eeq
where $\bp$ is the initial velocity and $\bx_0, v_0$ are integration constants.
\goodbreak

\item
Choosing instead $A_{--}= A_{-+} =0$  and $A_{+-}=A_{++}=1$ we get, by \eqref{BfromBJR}, the light-cone Milne form  of flat Minkowski spacetime  \eqref{Milne1} below,
\besub
\begin{align}
u_+&=U, \quad \bx_+ = \frac{\bX\;}{u_+},\quad
v_+ = V+\half  u_+\, \bx_+^2\,,
\\
P_{ij}^+(u_+)&=u_+\,\delta_{ij}\,,
\quad
a_{ij}^+(u_+)=u_+^2\,\delta_{ij} \,,
\quad
S_+^{ij} = -u_{+}^{-1}\,\delta^{ij}\,,
\label{PaSMilne}
\\[4pt]
ds^2_+&=2du_+dv_+ + u_+^2 d\bx_+^2.
\label{u2Minkowski}
\end{align}
\label{Milneform}
\esub
However these coordinates are valid only
in either of the half-spaces $U>0$ or $U<0$ but not for $U=0$, where the regularity condition $\det(a)\neq0$ is not satisfied.

Labeling the  translations by $c^i$ and the boosts by $b^j$ \emph{in Milne-BJR coordinates}, \eqref{aCarroll} yields the Milne-BJR implementation,
\beq
Y_{BJR}^+ = c^i \frac{\partial}{\partial x^i}
- b_i\left(\frac{1}{u_+}\frac{\partial}{\partial x_+^i} + x^i  \frac{\partial}{\partial v_+} \right)\,.
\label{MilneBJRtrb}
\eeq
Thus Milne translations are as in Minkowski space, but the Milne boosts look differently; the cast of parameters has also changed.

Again by \eqref{BJR0geo}, the geodesics in BJR-Milne coordinates are
\beq
\bx_+(u_+)=-\frac{1}{u_+}\,\bp +\bx_0, \qquad v_+= \frac{1}{2u_+} \bp^2+v_0\,.
\label{s1BJRgeo}
\eeq
Comparison with \eqref{s0BJRgeo} shows that the relations \eqref{+-BJRxyv} are duly satisfied.

These results are again consistent with those in sec.
\ref{SiklosSec} for $S=s_+=1, \, \mu_+=0$.

Pulling back to $B$ coordinates we get
\beq
Y = c_i\left( U\frac{\partial}{\partial X^i} - X^i  \frac{\partial}{\partial V} \right)
- b^i\frac{\partial}{\partial X^i}\,.
\label{BJRBtrb}
\eeq
Eqn. \eqref{addBrCaM} is  thus recovered, -- but with a different cast: a BJR-translation with parameter $c^i$ looks, in B coordinates, as a B-boost and a BJR${}_{+}$-boost with  $\bb$ looks as a B-translation with parameter $-\bb$,

In Brinkmann expression of the geodesics are obtained from  \eqref{BfromBJR}. The choice $s=s_-$ yields, as seen before, the standard expression \eqref{s0BJRgeo} with $X = x_-, V=v_-$.
But pulling back for $s_+$ we get, from \eqref{s1BJRgeo},
\beq
\bX_+(U_+)={U_+}\,\bx_0-\bp, \qquad V^+(U_+) =
\frac{1}{2U_+} \bp^2+v_0 -U_+\frac{1}{2} \bX_+(U_+)^2\,.
\label{s1Bgeo}
\eeq
consistently with \eqref{Bgeo+-}.

\end{itemize}

\goodbreak
\subsection{A Bianchi VII example: $C=1/2,\,\kappa=1$.}\label{C1k2subsub}

The parameters $C =\frac{1}{2}$  and $\kappa =1$ fall in the interior of the Bianchi VII range \eqref{BVIIrange}
and our previous investigations (those in subsec. \ref{IsoBJR} in particular)
apply. By \eqref{C<ks12} we have two real solutions,
$
s_{\pm} = \frac{1}{2}\pm\Delta.
$
The associated group parameters of Bianchi VII${}_h$, $h=k^{-2}$, are different, implying  different homogenous hypersurfaces.

\begin{figure}[ht]
\hskip-3mm
\includegraphics[scale=.3]{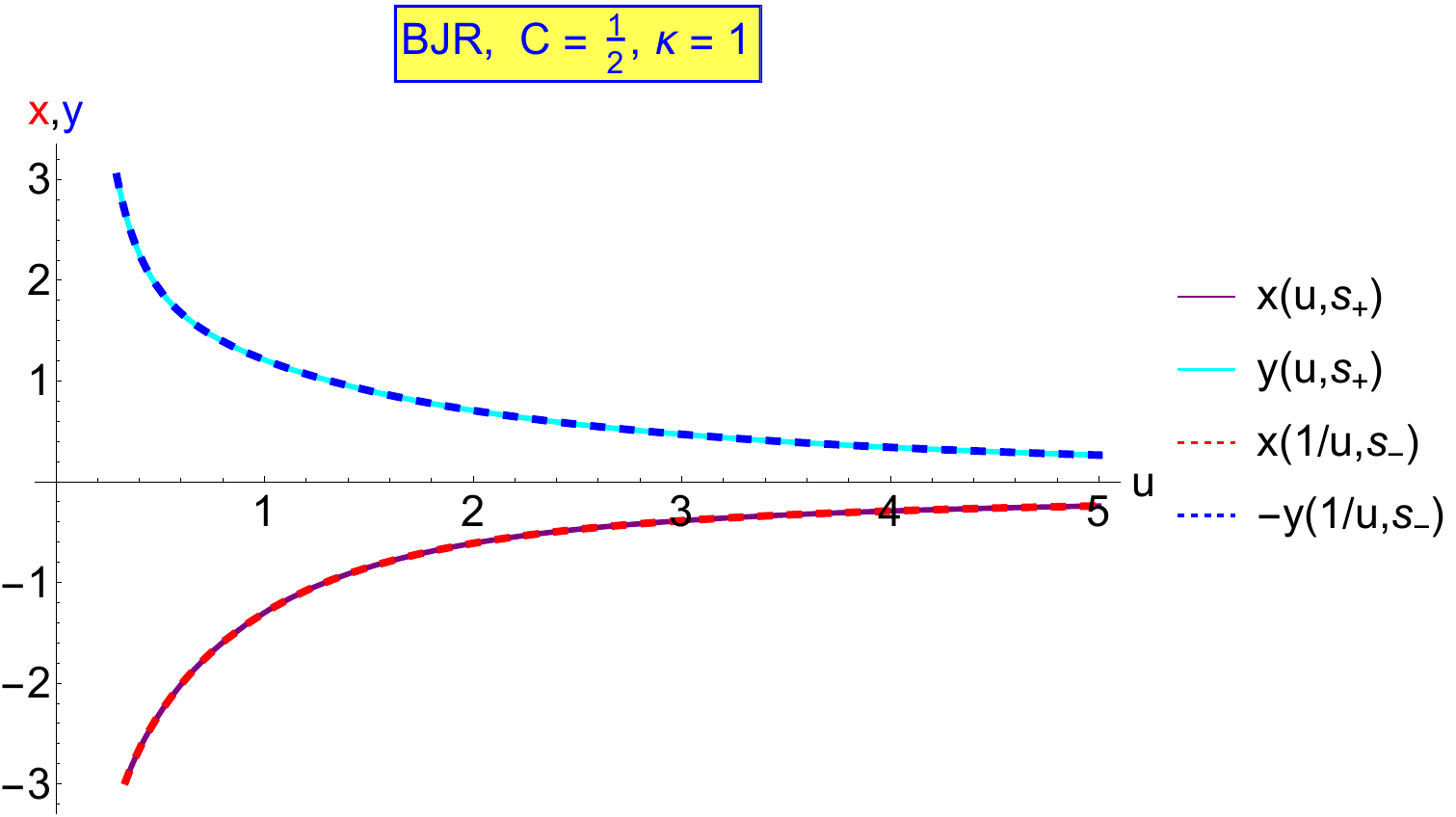}
\;\;
\includegraphics[scale=.27]{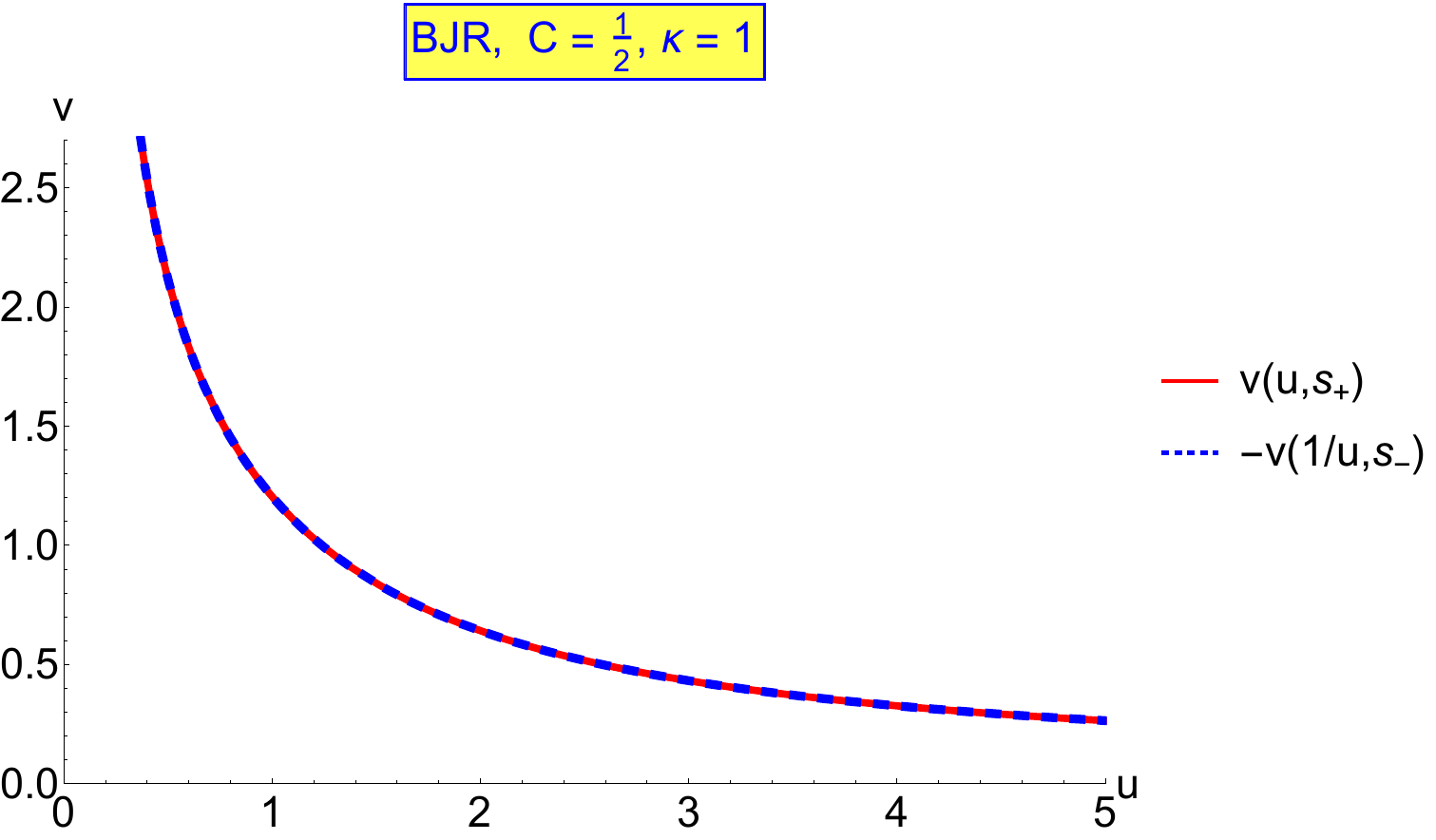}
\\{}\vskip-2mm
\hskip-12mm(i)\hskip74mm (ii)\\
\vskip-3mm
\caption{\textit{\small(i) Consistently with \eqref{+-BJRx} and \eqref{+-BJRy}, changing the parameter $u_+\equiv u$ into $u_-\equiv u^{-1}$ and multiplying by appropriate $\pm$ signs as in \eqref{+-uxyv} carries the transverse components of a geodesic associated with $s_+$ into those associated with $s_-$. (ii) For null geodesics
the vertical components $v(u_{\pm})$ match also, see \eqref{+-BJRv} .}
\label{fig2}
}
\end{figure}

Those ``translation-boost type'' isometries calculated in the two respective BJR coordinate systems associated with $s_+$ and $s_-$ and shown in fig.\ref{fig2} appear to be substantially different.
However they are carried into each other by following the rule \eqref{+-uxyv}, see fig.\ref{fig2}.
\goodbreak

\begin{figure}[ht]
\hskip-4mm
\includegraphics[scale=.3]{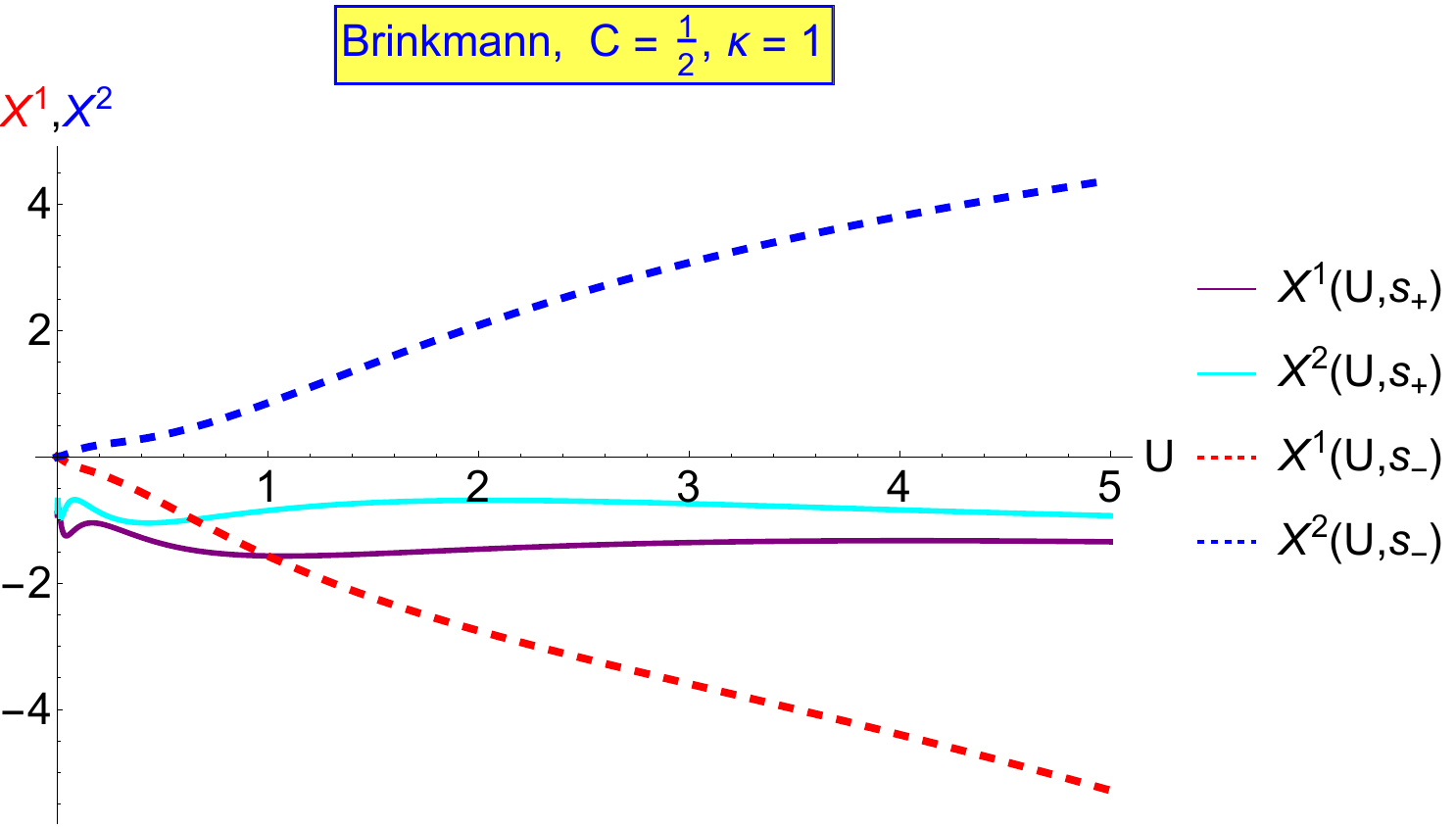}
\vskip-3mm
\caption{\textit{\small Pulling back to Brinkmann coordinates the transverse components of the BJR geodesics associated with $s_+$ or $s_-$ [or equivalently, the coefficients $\bbeta$ of the isometries] are different.}
\label{fig3}
}
\end{figure}

\begin{figure}[ht]
\hskip-3mm
\includegraphics[scale=.28]{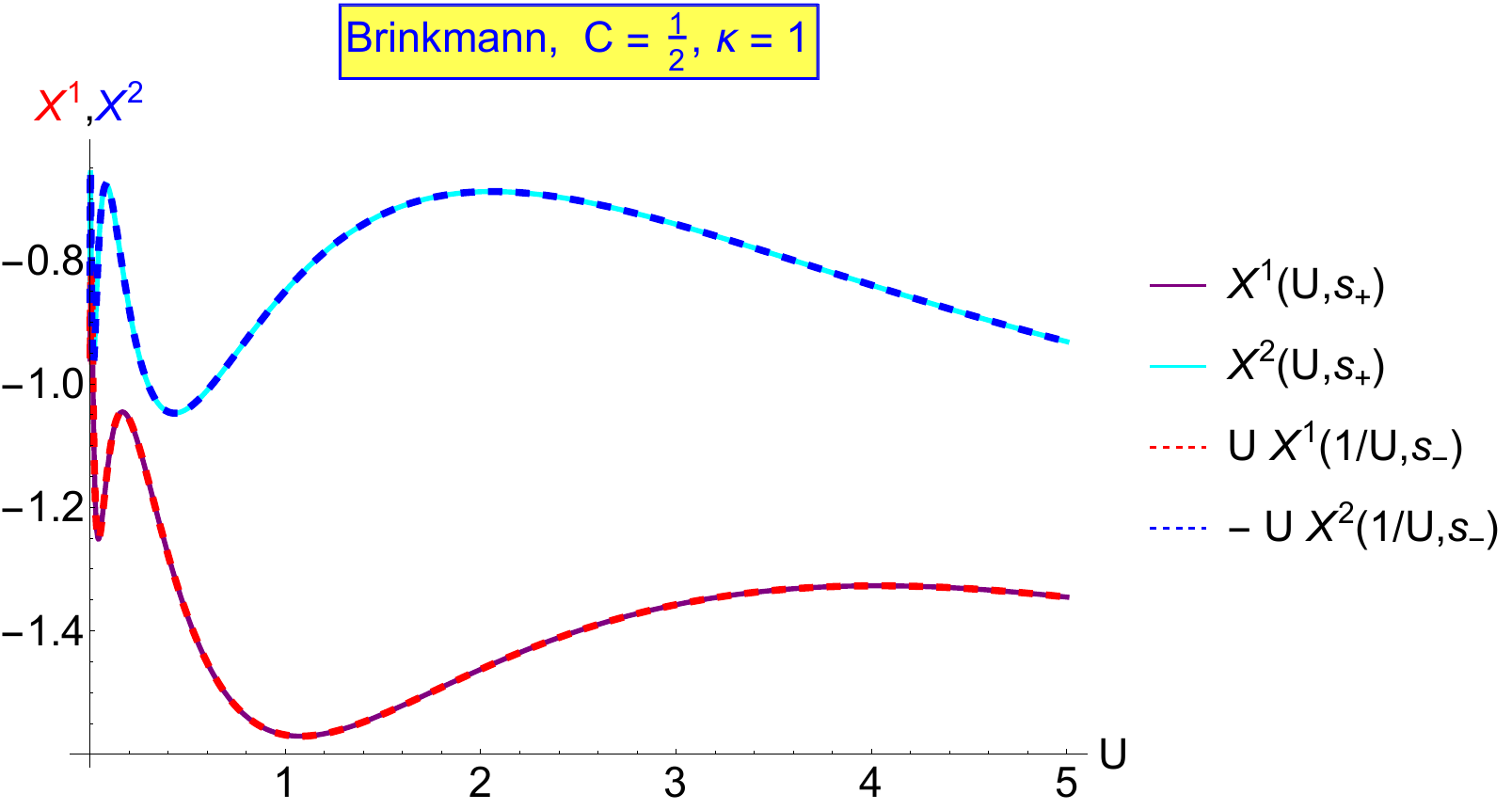}\;
\includegraphics[scale=.28]{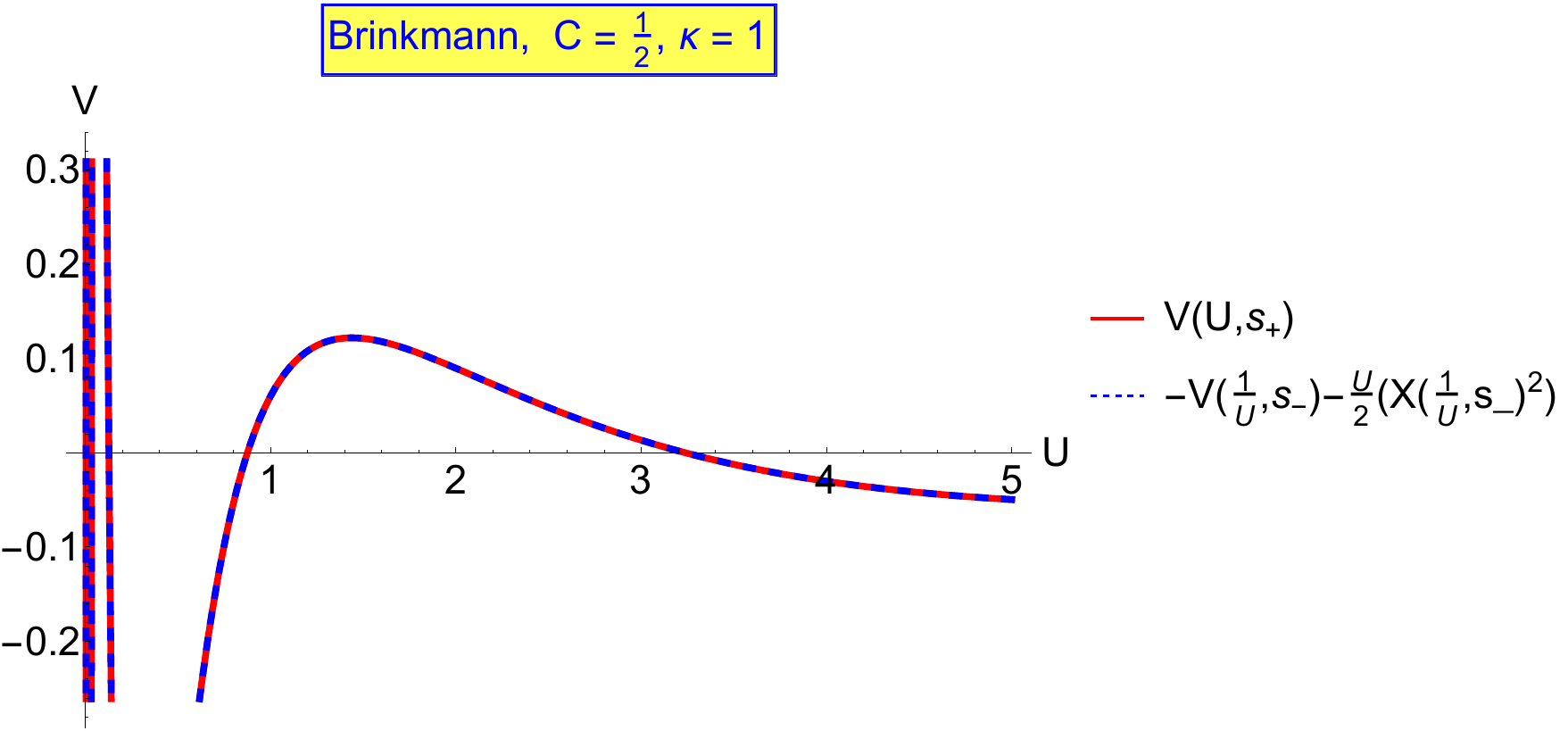}
\\
{}\hskip-25mm (i)\hskip70mm (ii) \\
\vskip-3mm
\caption{\textit{\small The transverse projections of the geodesics in Brinkmann coordinates are taken into each other under \eqref{Bgeo+-}.}
\label{fig4}
}
\end{figure}

Similarly, solving (either numerically or by pulling back from BJR to B coordinates by \eqref{BfromBJR}) the transverse equations \eqref{ABXeq} in Brinkmann coordinates yields differently looking geodesics, see fig.\ref{fig3}.
However combining with the inversion $U \to U^{-1}$
and changing  signs appropriately yields identical geodesics, consistently with \eqref{Bgeo+-}, see
fig.\ref{fig4}.

The  Killing vectors are $Y_V=\p_V$ and $Y_\kappa$ in \eqref{kUVboost}, augmented with those 4 translation-boost type ones  in  eqns. \eqref{addBrCa} which can be calculated both  numerically or analytically using 
 Appendix B.

\subsection{A Bianchi VII example: $C=\kappa$}\label{C=ksubsub}

For the value $C=\kappa$ which lies at the edge of the  range \eqref{BVIIrange}, eqn \eqref{4s} yields 4 solutions,
$s = \half$ \;(double root)
and
$s = \half(1 \pm\sqrt{1-4C^2}\big)\,.$
Although the latter two are real when $C=\kappa \leq 1/2$ they should nevertheless be discarded because of the
Bianchi VII requirement $C=\kappa > 1/2$ in \eqref{BVIIrange}
leaving us with just one (double) real solution,  $s=\half$;
 \eqref{kbmualpha} yields the auxiliary parameters.
The Souriau matrix is,
\beq\begin{array}{cll}
S^{11} &=& \,\frac{2|\kappa|}{\sqrt{4\kappa^2-1}} \,\ln u + \frac{1}{4\kappa^2-1} \,\sin(\sqrt{4\kappa^2-1}\,\ln u) \,, \qquad
\\[6pt]
S^{12}= S^{21} &=&  \frac{1}{4\kappa^2-1}\,\cos (\sqrt{4\kappa^2-1}\,\ln u) \,,
\\[6pt]
S^{22} &=&  \frac{2|\kappa|}{\sqrt{4\kappa^2-1}}\,\ln u - \frac{1}{4\kappa^2-1} \,\sin (\sqrt{4\kappa^2-1}\,\ln u) \,.
\end{array}
\label{Ck1Smatrices}
\eeq
\begin{figure}[ht]
\includegraphics[scale=.25]{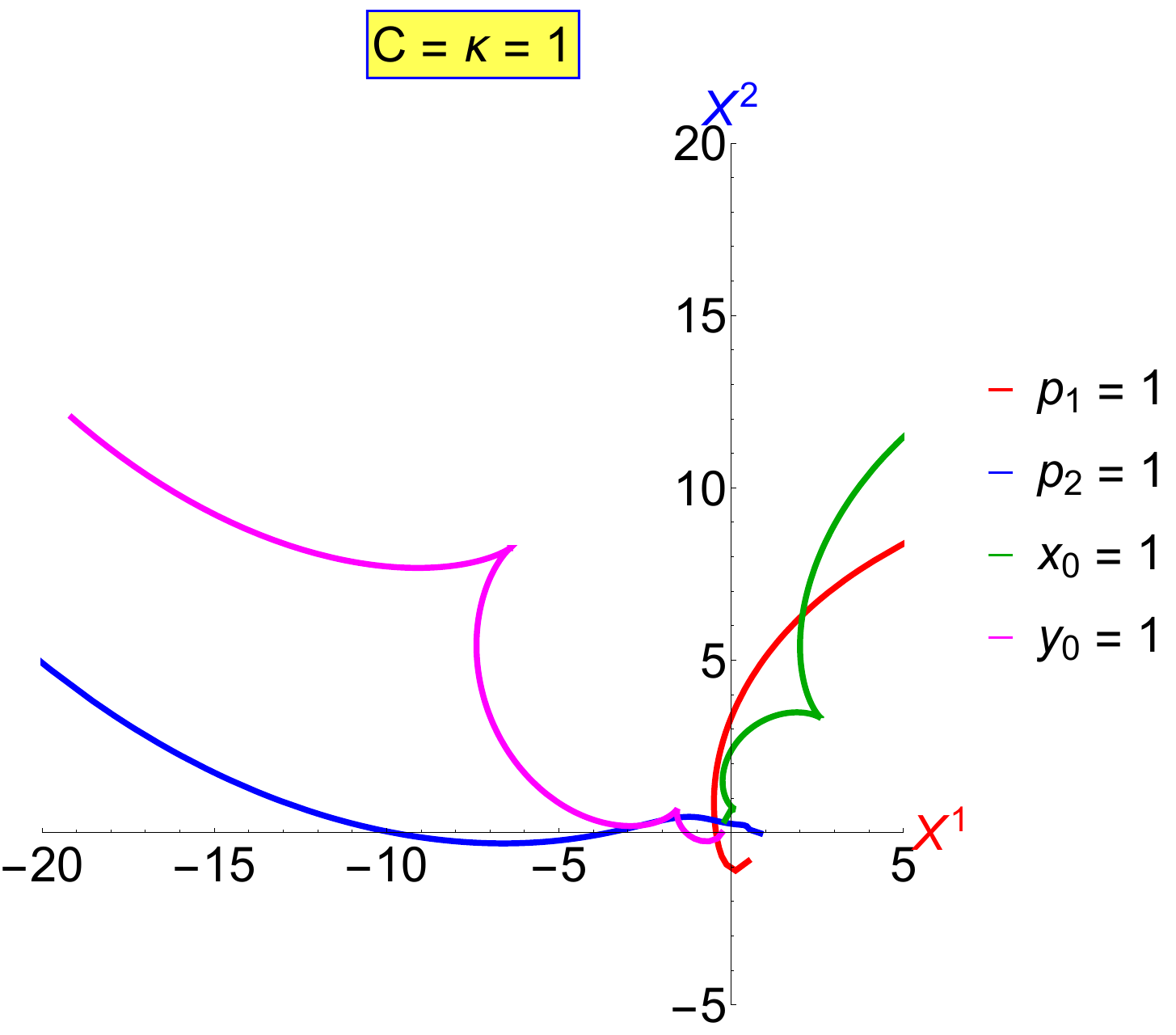}
\;
\includegraphics[scale=.25]{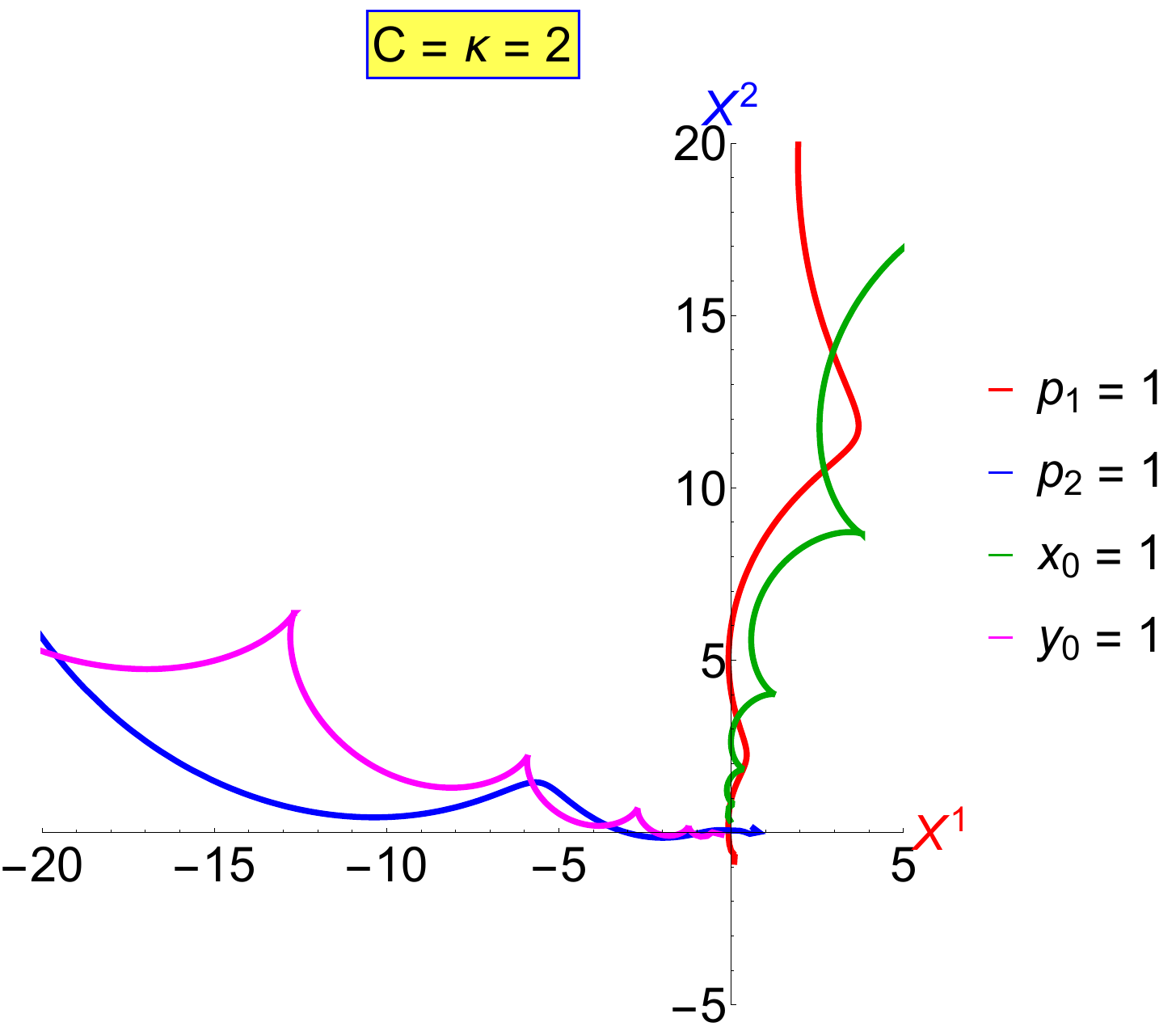}
\\
\includegraphics[scale=.25]{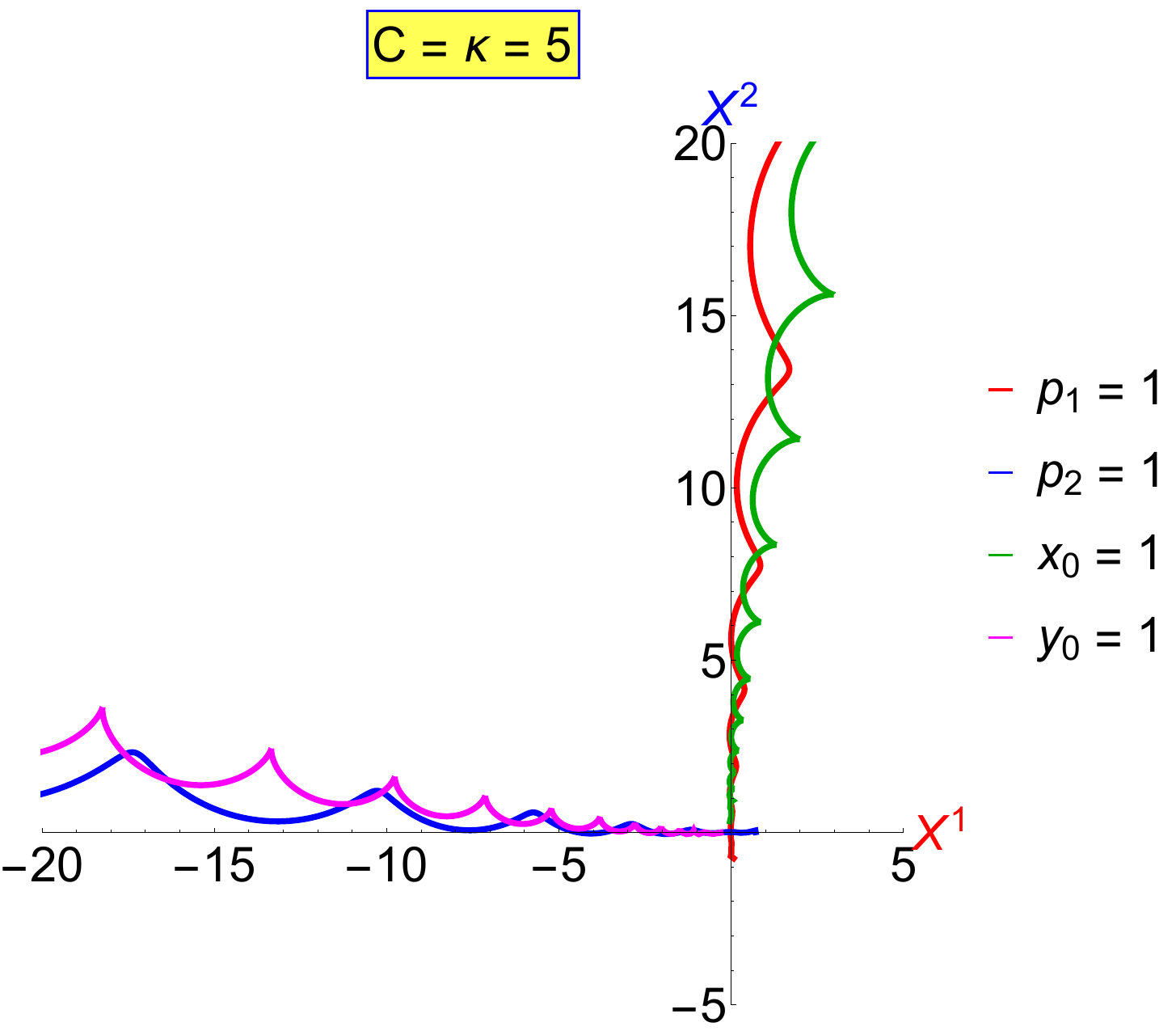}
\;
\includegraphics[scale=.25]{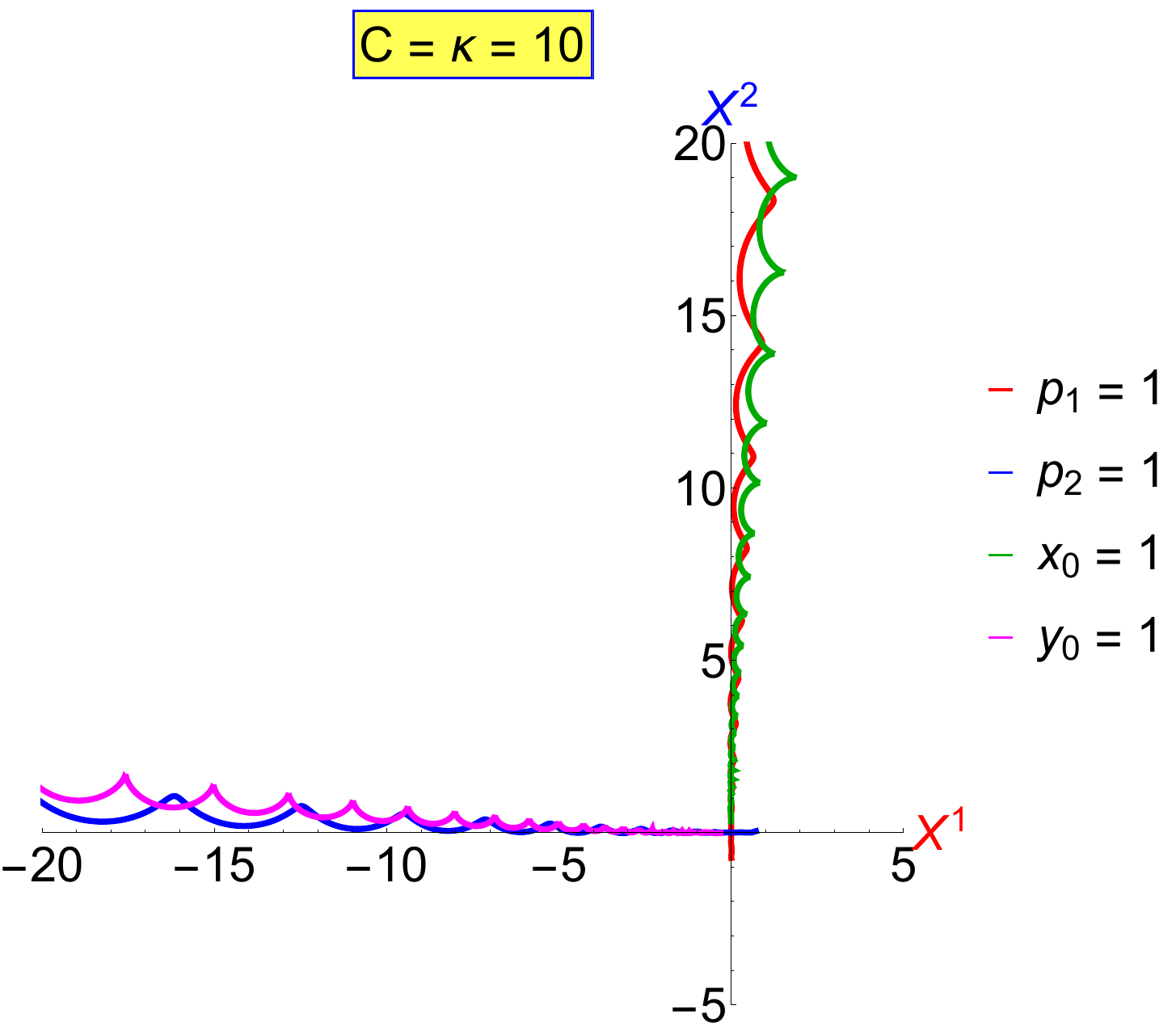}
\vskip-4mm
\caption{\textit{\small Transverse components of 4-parameter families of ``translation-boost type'' isometries/geodesics
 for Bianchi-VII-type Lukash waves.
 When $C=\kappa\to \infty$ they  manifestly ``flatten out.}
\label{fig5}
}
\end{figure}
Then \eqref{addBrCa} (resp. \eqref{ABXeq} -
\eqref{Vpp}) yield, for each fixed value of $C=\kappa~>~\2$,
 a 4-parameter family of Killing vectors (resp. geodesics),  whose transverse components are plotted in fig.\ref{fig5}. Unfolding  them to ``time'' $U$  would indicate that starting from a small initial value $U_0>0$, they  extend to $U\to\infty$.

To explain the curious ``flattening'',  we could argue intuitively that for increasingly high frequency $\kappa$ the wave has no ``time" to push out the geodesic before pulling it back again. This can also be confirmed analytically. Letting  $\kappa \to \infty$ the P matrix  \eqref{complexP} resp. the  Souriau matrix \eqref{Ck1Smatrices}
 become asymptotically linearly polarized (off-diagonal) and  resp. diagonal.
 Inserting their asymptotic values
into eqn \eqref{Y1} \eqref{Y2} \eqref{Y3} \eqref{Y4} of Appendix B yields the asymptotic behavior
\beq
Y_1 \approx \sqrt{U}\,\p_2 ,\quad
Y_2 \approx -\sqrt{U}\,\p_1 ,\quad
Y_3 \approx  (\sqrt{U}\,\ln u)\,\p_2, \quad
Y_4 \approx  (-\sqrt{U}\,\ln u)\,\p_1\,.
\label{asymptY}
\eeq
Equivalently, the transverse components of the geodesics are
 asymptotically
\beq
X^1
\approx -\sqrt{U} (p_y\,\ln U + y_0),\qquad
X^2
\approx \sqrt{U} ( p_x\,\ln U + x_0)
\label{asymptgeo}
\eeq
consistently with fig.\ref{fig5}.

\section{Global Considerations} \label{GlobalSec}
%
%
\kikezd{The Milne Region}.
In order to get an insight into the global structure,  it may be helpful to consider first the flat limit  when no wave is present, i.e., when  $C=0$
in Brinkmann coordinates which is
Minkowski spacetime, with $-\infty < U < \infty $, $-\infty < V < \infty $.
If $ U=0$ we have a null-hyperplane separating Minkowski spacetime in
two halves, one half to its future, $U>0$, and one to its past $U<0$.
If $C\ne 0$ $U=0$ then is singular and we can with no loss of generality restrict $U$ to positive values.
Setting
$
U= u,\,  \bX=u \bx,\,  V= v - \half{u}\,\bx^2
$
allows us to the metric in the form,
\beq
ds^2 = 2 dU dV + d\bX^2 = 2 dudv + u^2 d\bx^2,
\label{Milneuv}
\eeq
cf. \eqref{u2Minkowski}.
The coordinate transformation is valid
only if $U=u>0$ or $U=u<0$ and is singular on the null hyperplane $u=U=0$.
The further coordinate change (valid only if $u>0,\, v<0$)
\beq
u =  t\exp(-z) \,,\qquad v = - \half t\exp(z)
\label{coordtrans2}
\eeq
transforms the metric to
\beq
ds^2 = - dt^2 + t^2 \Bigl\{dz^2 + e^{-2z}(dx^2 + dy^2)\Bigr\} \,.
\label{Milne1}
\eeq
which is the \emph{Milne form} of the flat Minkowski metric.
Let us call this region II, since it is to the future of the region I which is defined as $u>0,\, v>0$.

Since the 3-metric in the braces   is of constant negative curvature the spacetime metric is of Friedmann-Lema\^{\i}tre form with $t$ playing the role of cosmic
time and $x,y,z$ being comoving coordinates.
The 3-metric in braces in  (\ref{Milne1}) is that of the upper-half space model of three-dimensional
hyperbolic space $H^3$, but as such is not  global.

The isometry  group of
hyperbolic space is $\SO(3,1)$ and (\ref{Milne1}) makes manifest the Bianchi VII subgroup, together with an additional $\SO(2)$, the so-called LRS action.

\goodbreak
\kikezd{The Rindler Region}.
 %
Having dealt with region II we now return to consideration
of region I, that is $ v>0$ .
We replace (\ref{coordtrans2}) by reversing the roles played by of $z$ and $t$,
\beq
u=  \tilde z\, \exp({\tilde t})   \,,\qquad v=  \half   \tilde z \, \exp(-\tilde t) \label{Rindler0}
\eeq
and find the \emph{Rindler form},
\beq
ds ^2 = d \tilde z^2 + \tilde z^2
\Bigl\{-  d\tilde t^2 + e^{2\tilde t} (dx^2 + dy^2) \Bigr\} \,.
\label{Rinder1}
\eeq
The metric in braces is Lorentzian and is  that of one half of three-dimensional
de Sitter spacetime $dS_3$. It has constant curvature. In terms of the Carter-Penrose diagram the metric is
\beq
2 du dv = -(dx^0)^2 + (dx^3)^2 =  - \tilde z ^2 d \tilde t ^2 + d \tilde z ^2
\eeq
and is valid inside the so called right hand Rindler Wedge $ x^3 > |x^0|$.
Since
\beq
x^0 = \tilde z \sinh \tilde t \,,  \qquad x^3 = \tilde z \cosh \tilde t
\eeq
the  timelike curves $ \tilde z = {\rm constant} >0 $
are hyperbolae and the orbits  of Lorentz  boosts about the origin.
As a consequence they may  consider the world lines of Rindler observers  having  constant
acceleration. The null line  $ \tilde t = \tilde z$
is their  future horizon and the null line  $ \tilde t = -\tilde z$ their past horizon.
This should be contrasted with the Milne region II.
In that case
\beq
ds^2 = 2 du dv =  (dx^0)^2 - (dx^3)^2 = -d t ^2 + t^2 d z ^2
\eeq
whence \vspace{-3mm}
\beq
x^0 = t  \sinh  z \,, \qquad  x^3 = t  \cosh z \,.
\eeq
The curves $z= {\rm constant} $ are timelike straight lines
through the origin which may be considered the worldlines of Milne's cosmological observers.


Now we turn to  Lukash plane waves. The discussion
  closely parallels that for Minkowski spacetime given above
  but using the material developed
  earlier. We restrict the Lukash metric to positive $U$.
  Recall that  in  section \ref{BianchiformSec} we introduced
 the coordinates $t$ and $z$ as in \eqref{tzdefinition} and cast the metric in  Bianchi $VII_h$ form as in  equations \eqref{gij}.
 This was valid for $u>0,\, v<0$. The orbits of the Bianchi $VII_h$
 are spacelike. This is sufficient to cover the Milne type region.
In order to cover the  Rindler type region $u>0,\, v>0$
 we cannot use  \eqref{tzdefinition} but rather
introduce coordinates $\tilde t, \tilde z$ by
\beq
u=  \tilde t  e^{-\frac{\tilde z}{2b} } \,,\quad
v= \half\tilde t e^{\frac{\tilde z}{2b}}
\qquad\text{\small in terms of which}\qquad
2dudv = d {\tilde t}^2 - {\tilde t}^2\frac{d \tilde z^2 }{4b^2}
\label{V.14}\eeq
and so  $\p_{\tilde t}$ is spacelike and  $\p_{\tilde z}$ is timelike.
Now \eqref{Bianchiform} is replaced by
\beq
ds ^2 = (d \tilde t)^2 + g_{ij}(\tilde t)\lambda^i\lambda^j \,,
\label{V.15}\eeq
where the invariant one-forms depend on $ \tilde z, x , y$
and the 3-metric   $g_{ij} ( \tilde t) $ now has Lorentzian signature $++-$ .

As remarked previously, the  pp-wave metric  is singular at $u=0$, and therefore we restrict  our exploration  to $u>0$.
In fact, he singularity is of a type called ``non-scalar''
because no scalar polynomial  formed from the Riemann tensor
blows up. In fact for pp-waves all scalar polynomials  formed from the Riemann tensor vanish identically.

The situation is illustrated by  figure 1, p.~406
of \cite{Siklos81} and  figure 1  on p.~256 of \cite{Siklos3}.

\section{Conclusion}\label{Conc}

In this paper we have examined in detail
 the Lukash plane gravitational wave. Our aim  has been to give a self-contained account of its geometry and  global structure and to relate it to
spatially homogeneous cosmological models.

pp waves admit a generic five dimensional isometry group  which acts within the wave fronts \cite{BoPiRo,exactsol,LL,Sou73,Carroll4GW,Carrollvs}.

The Lukash wave metric has (as do CPP waves) an additional $6$th isometry
 \cite{exactsol, Carroll4GW, CPP, Conf4GW}, given in
eqn \eqref{kUVboost} (in Brinkmann) or in \eqref{Y6} in (BJR). The extra generator takes one out of the wave fronts  and the metric becomes  homogeneous.

This additional generator actually belongs to a three dimensional Bianchi $VII_h$ type subgroup which acts transitively on three dimensional orbits and leads to an intimate connection between gravitational waves and  spatially homogeneous cosmological models and thence to the theory of Killing horizons.
 In BJR coordinates the Bianchi group
 ${G}_3$ consists of two transverse translations plus the VSR-type triple combination \cite{GibbonsVSR},
 cf. \eqref{BianchiVIIKvs}- \eqref{Y6}.

In section \ref{GlobalSec} we provide a global picture of the spacetime. The gravitational wave emanates from a singular wave front and splits into two regions which we have dubbed  of \emph{Milne type} and of \emph{Rindler type}, divided  by a Killing horizon.

In the  Milne region the orbits of the Bianchi
group are spacelike and the spacetime resembles an anisotropic deformation of Milne's cosmological model. In the Rindler region the orbits are timelike and the spacetime resembles an anisotropic deformation of the Rindler wedge.
The  involution \eqref{univ} interchanges the singularity at $u=0$ with the conformal infinity at $u=\infty$
as observed in  sects \ref{Multis} and  \ref{GlobalSec}
and reserved for further investigations.

\begin{acknowledgments}
This work was partially supported by the National Key Research and Development Program of China (No. 2016YFE0130800) and the National Natural Science Foundation of China (Grant No. 11975320). During the final phase of preparations ME was supported by TUBITAK grant 117F376.
\end{acknowledgments}


\bigskip
\renewcommand{\theequation}{A\thesection.\arabic{equation}}
\appendix
{\bf Appendix A: The Bianchi  $VII_h$ type Group}\label{BVIIsec}

\renewcommand{\theequation}{A\thesection.\arabic{equation}}
\appendix{\appendix}{\setcounter{equation}{0}}

Here we collect some relevant facts about the Bianchi group $VII_h$ \cite{exactsol,EllisMacCallum,MacCallum}. The latter has a matrix representation \vspace{-4mm}
\beq
g(\lambda, \eta)  = {  \left(
\begin{array}{ccc}
e^{i(1-ic)\lambda }&0& \eta\\
0& 1 & \lambda \\
0&0&1 \\
\end{array}
\right)}
\eeq
where $\lambda \in \Bbb{R}$ and $\eta \in \Bbb{C}$.
To obtain a  real representation one  sets
$
i= {  \left(
\begin{array}{cr}
0& -1\\
1& 0   \\
\end{array}
\label{imatrix}
\right)
}
$
in the first column  and  replaces $\eta$ by
$
{  \left (\begin{array}{c}
\alpha\\ \beta \\ \end{array} \right)}
$
with $\alpha, \beta \in \Bbb{R} $.
One may verify that
\beq
g(\lambda_1,\eta_1) g(\lambda_2,\eta_2) = g(\lambda_1 +\lambda_2,
e^{i(1-ic)\lambda_1} \eta_2 + \eta_1)\,.
\label{left}
\eeq
Infinitesimally
\besub
\begin{align}
g &=  {\left(
\begin{array}{ccc}
1&0& 0\\
0& 1 & 0 \\
0&0&1 \\
\end{array}\right)
+ \lambda \left (\begin{array}{ccc}
i(1-ic)&0& 0\\
0& 0 & 0 \\
0&0&0 \\
\end{array}  \right ) + \eta \left (\begin{array}{ccc}
0&0& 1\\
0&0& 0 \\
0&0&0 \\
\end{array} \right) + \dots }
= 1 + \alpha \be _1 + \beta \be_2 + \lambda \be _3 + \dots
\end{align}
\esub
where
\beq
\be_1= {  \left(
\begin{array}{cccc}
0 & 0& 0 &   1\\
0& 0 & 0 &0\\
0&0&0 &0 \\
0&0&0&0
\end{array}
\right)\,,  \qquad \be_2= \left(
\begin{array}{cccc}
0& 0& 0& 0\\
0& 0 & 0 &1 \\
0&0 &0 &0 \\
0&0&0&0
\end{array}
\right)\,, \qquad
\be_3= \left(
\begin{array}{cccc}
c& -1&  0 & 0\\
1& c  &0 &0 \\
0& 0 & 0 &  1 \\
0&0&0 &0
\end{array}
\right)\,.}
\eeq
The $\be_1,\be_2,\be _3 $ form a basis 
of the Lie algebra $\mathfrak{vii}_h$
with commutation relations
\beq
[\be_3,\be_1 ]= c\, \be_1 + \be_2 \,,\qquad  [\be_3,\be_2 ]= c \,\be_2 - \be_1  \,.
\eeq \vspace{-2mm}
Thus we get
\beq
g^{-1} dg = \be_1 \lambda^1 + \be_2 \lambda^2 + e_3 \lambda^3\,.
\eeq
 $ \lambda^i$ is a basis of left-invariant one-forms given by
\beqa
\lambda^1 &=&   e^{-c \gamma} (\cos \gamma d\alpha  + \sin\gamma d \beta)
\nn\\
\lambda^2 &=&   e^{-c\gamma}  (\cos \gamma d\beta - \sin \gamma d\alpha)
\label{oneforms}
\\
\lambda^3 &=& d \gamma \nn
\eeqa
and the non-trivial Maurer-Cartan relations are
\beq
d \lambda^1 =-c \lambda ^3 \wedge \lambda ^1  +\lambda ^3 \wedge \lambda ^2
\,,\qquad   d \lambda ^2 =-c \lambda ^3 \wedge \lambda ^2  -\lambda ^3 \wedge
\lambda ^1\,.
\eeq

In terms of structure constants: if
\beq
[\be_i,\be_j]= C_i\, ^k \,_j \be_k \label{Lie}
\eeq
then
$d\lambda^k = - \half C_i\, ^k \,_j \lambda^i \wedge \lambda^j \,,
$
where \vspace{-2mm}
\beqa
 C_3\,^1\,_1 = -  C_1\,^1\,_3 = c \,,&\qquad&  C_3\,^2\,_1 = -C_1\,^2\,_3 =  1 \nn\\
 C_3\,^2\,_2 = -  C_2\,^2\,_3 = c \,,&\qquad&  C_3\,^1\,_2 = -C_2\,^1\,_3  =-1 \,.
\label{CRS}
\eeqa
The vector fields generating left actions on $VII_h$ are
obtained by taking $\lambda_1$ and $\eta_1$  infinitesimal. From \eqref{left} we find
$
\delta \eta_2 = \eta_1\,, \, \delta \lambda = \lambda_1
 + i(1-ic) \eta_i
$
which correspond to:
\beq
  R_1 = \p_\alpha \,, \qquad R_2=\p_ \beta  \,, \qquad R_3=
  \p_\lambda + \alpha  \p_\beta -\beta  \p_\alpha
  +c(\alpha \p_\alpha  +\beta  \p_\beta ) \,,
\eeq
\vskip-2mm
  whence \vspace{-2mm}
\beq
[R_3, R_1]= -c R_1 - R_2 \,,\qquad  [R_3, R_2]= -c R_2 + R_1
\,.
\eeq
  In other words,
  \beq
[R_i, R_j] = - C_i\, ^k \,_j R_k \,.
\eeq

Cartan's formula
$
\cL_V \omega = d (i_X \omega) + i_X d\omega
$
with  $V=R_i$ and $\omega= \lambda^j $ shows that  the one-forms $\lambda^j$ are left-invariant,
$
\cL_{R_i} \lambda ^j =0\,.
$
From \eqref{CRS} we deduce that
\beq
C_i\,^k\,_k= (0,0,2c)
\eeq
which confirms that the Bianchi $VII_h$ is not
\emph{non-unimodular}, that
is, the adjoint representation is not volume-preserving, known in the cosmology literature as \emph{class B}.

To make contact with \eqref{BJRmetric} it is helpful to note that
\besub
\begin{align}
\bigl( \lambda^1 \bigr)^2 + \bigl(\lambda^2 \bigr)^2 &=  e^{-2c \lambda} (d \alpha^2 + d \beta ^2 ) \label{sum}\\
\bigl( \lambda^1 \bigr)^2 - \bigl(\lambda^2 \bigr)^2 &=  e^{-2c \lambda}
(\cos (2\lambda) (d\alpha^2 - d\beta^2) + 2\sin (2\lambda) d\alpha d \beta) \label{difference}
\\
2\lambda^1   \lambda ^2 &= e^{-2c\lambda}(\cos (2\lambda) 2 d \beta d \alpha - \sin (2 \lambda)
( d \alpha^2 - d \beta^2 )) \,
\label{product}
\end{align}
\esub\vskip-4mm
so that \vspace{-2mm}
\beqa
&&k(\lambda^1)^2 + k(\lambda^2)^2 + (\lambda^1)^2-(\lambda^2)^2 =
\nn
\\
&& \quad
e^{-2c\lambda} \Bigr\{(k+\cos 2\lambda) d \alpha^2
+ 2 \sin 2\lambda d\alpha d\beta + (k-\cos 2\lambda) d\beta ^2   \Bigr\}\,.
\label{2metric}
\eeqa

\bigskip
{\appendix}{\setcounter{equation}{0}}
\renewcommand{\theequation}{B\thesection.\arabic{equation}}
{\bf Appendix B:
 ``Translation-boost-type'' isometries}\label{Btrbsec}

%

In the Lukash case the $P$-matrix is known in terms of the Siklos' parameters \eqref{4s}-\eqref{kbmualpha} and
spelling out eqn \eqref{addBrCa} yields those 4 ``translation-boost-type'' isometries in Brinkmann coordinates,
\begin{itemize}
\item 2 ``translations''
\beq
Y_1 = P_{11} \partial_1 + P_{21} \partial_2 - (P_{11}' X^1 + P_{21}' X^2) \partial_V\,,
\label{Y1}
\eeq\vskip-4mm
where
\begin{subequations}
\begin{align}
P_{11} &= \frac{U^s}{2} \left[ e^{i\alpha} \Big( \cosh(\mu/2) U^{-i(\kappa - b)} - \sinh(\mu/2) U^{-i(\kappa +b)}  \Big) + c.c.   \right]\, \\
P_{21} &= \frac{U^s}{2} \left[ -i e^{i\alpha} \Big( \cosh(\mu/2) U^{-i(\kappa - b)} - \sinh(\mu/2) U^{-i(\kappa +b)}  \Big) + c.c.   \right]\,
\end{align}
\label{addBCex1}
\end{subequations}
\noindent
and
\beq
Y_2 = P_{12} \partial_1 + P_{22} \partial_2 - (P_{12}' X^1 + P_{22}' X^2) \partial_V\,,
\label{Y2}
\eeq
\vskip-3mm
 where
 \begin{subequations}
\begin{align}
P_{12} &= \frac{U^s}{2} \left[ i e^{i\alpha} \Big( \cosh(\mu/2) U^{-i(\kappa - b)} + \sinh(\mu/2) U^{-i(\kappa +b)}  \Big) + c.c.   \right]\, \\
P_{22} &= \frac{U^s}{2} \left[  e^{i\alpha} \Big( \cosh(\mu/2) U^{-i(\kappa - b)} + \sinh(\mu/2) U^{-i(\kappa +b)}  \Big) + c.c.   \right]\,
\end{align}
\label{addBCex2}
\end{subequations}
\vspace{-8mm}
\item
These are completed with 2 ``boosts"
\beq
Y_3 = (S^{11}P_{11} + S^{12}P_{12})\partial_1 + (S^{11}P_{21} + S^{12}P_{22})\partial_2 -  (S^{1k}P_{jk})' X^j \partial_V\,,
\label{Y3}
\eeq
whose $\partial_1$ resp. $\partial_2$ components are
\beqa
 S^{11}P_{11} + S^{12}P_{12} &=& \frac{U^{1-s}}{2} \left[ e^{i\alpha} U^{-i(\kappa - b)} \frac{\cosh(\mu/2) (1-2s + 2i\kappa)}{(1-2s)(1-2s+ 2ib)} + c.c  \right]\,, \nonumber \qquad \qquad
 \\[6pt]
 &+& \frac{U^{1-s}}{2} \left[ e^{i\alpha} U^{-i(\kappa + b)} \frac{\sinh(\mu/2) (1-2s + 2i\kappa)}{(1-2s)(1-2s - 2ib)} + c.c  \right]\,,
 \eeqa
\beqa
 S^{11}P_{21} + S^{12}P_{22} &=& \frac{U^{1-s}}{2} \left[ -i e^{i\alpha} U^{-i(\kappa - b)} \frac{\cosh(\mu/2) (1-2s + 2i\kappa)}{(1-2s)(1-2s+ 2ib)} + c.c  \right]\,, \nonumber \qquad \qquad
 \\[6pt]
 &+& \frac{U^{1-s}}{2} \left[ -i e^{i\alpha} U^{-i(\kappa + b)} \frac{\sinh(\mu/2) (1-2s + 2i\kappa)}{(1-2s)(1-2s - 2ib)} + c.c  \right]\,.
 \eeqa
and by
 \beq
 Y_4 = (S^{21}P_{11} + S^{22}P_{12})\partial_1 + (S^{21}P_{21} + S^{22}P_{22})\partial_2 -  (S^{2k}P_{jk})' X^j \partial_V\,,
 \label{Y4}
 \eeq
whose $\partial_1$ resp. $\partial_2$ components are
\beqa
S^{21}P_{11} + S^{22}P_{12}&=& \frac{U^{1-s}}{2} \left[ i e^{i\alpha} U^{-i(\kappa - b)} \frac{\cosh(\mu/2) (1-2s + 2i\kappa)}{(1-2s)(1-2s+ 2ib)} + c.c  \right]\, \nonumber \qquad \qquad
\\[6pt]
 &+& \frac{U^{1-s}}{2} \left[ -i e^{i\alpha} U^{-i(\kappa + b)} \frac{\sinh(\mu/2) (1-2s + 2i\kappa)}{(1-2s)(1-2s - 2ib)} + c.c  \right]\,,
\label{C9}
\eeqa
\beqa
 S^{21}P_{21} + S^{22}P_{22} &=& \frac{U^{1-s}}{2} \left[ e^{i\alpha} U^{-i(\kappa - b)} \frac{\cosh(\mu/2) (1-2s + 2i\kappa)}{(1-2s)(1-2s+ 2ib)} + c.c  \right]\, \nonumber \qquad \qquad
 \\[6pt]
 &+& \frac{U^{1-s}}{2} \left[ - e^{i\alpha} U^{-i(\kappa + b)} \frac{\sinh(\mu/2) (1-2s + 2i\kappa)}{(1-2s)(1-2s - 2ib)} + c.c  \right]\,.
 \label{C10}
\eeqa

\end{itemize}

\end{document}